\documentstyle[aps,epsfig]{revtex}
\newcommand{\lsim}{\mathrel{\rlap{\lower4pt\hbox{\hskip0pt$\sim$}}
\raise1pt\hbox{$<$}}}
\begin{document}

\title{\large\bf  BOUND STATES IN QUANTUM FIELD THEORY, SCALAR FIELDS}
\author{G.V. Efimov \vspace*{0.2\baselineskip}}

 \address{Bogoliubov Laboratory of Theoretical Physics,
 \\Joint Institute for Nuclear Research, 141980 Dubna,
 Russia\vspace*{0.2\baselineskip} }
\maketitle
\begin{abstract}

The main aim of this paper is to demonstrate the method
called "the Bosonization of Nonlocal Currents" (BNC), used
for calculations of bound states in a quark model, within
the simplest relativistic quantum field model of two scalar
fields with the Yukawa type interaction:
$$ L=\Phi^+(\Box-M^2)\Phi+{1\over2}\phi(\Box-m^2)\phi
+g\phi\Phi^+\Phi.$$
A second aim is to clarify the relation between  BNC
and two widely used methods, employed in recent particle
physics to calculate bound states of interacting particles,
based on the nonrelativistic Schr\"{o}dinger equation
(the S-method), and the relativistic Bethe-Salpeter equation
(the BS-method), and to determine the conditions on parameters
of a quantum field model dictating a definite method to be
applied.

It is shown that all these methods can be applied only in
the weak coupling regime (the effective dimensionless coupling
constant $\lambda$ should be less than 1). The basic parameter
separating the relativistic and nonrelativistic pictures is
$\xi={m\over M}$, namely, $\xi\ll1$ with $\lambda\ll1$ leads
to the potential picture, i.e., the bound state is described by
the nonrelativistic Schr\"{o}dinger equation. For $\xi\geq1$
and $\lambda\leq1$ the Schr\"{o}dinger potential picture is
not valid and the Bethe-Salpeter equation or the BNC should be
used, where the BNC method has a slightly bit wider region of
applicability.

\vspace{0.3cm}
\noindent
PACS number(s): 03.65.Ge, 03.65.Db, 03.70.+k, 11.10.St, 11.15.Tk

\end{abstract}

\section{Introduction.}

Great efforts have been made to understand how bound states
arise in the formalism of quantum field theory and to work out
effective methods to calculate all characteristics of these
bound states, especially their masses.
Unfortunately, we observe that there is no well-defined
unique method, like the Schr\"{o}dinger equation in
nonrelativistic quantum mechanics, which can be used practically
for any problems of nonrelativistic quantum physics. We
can conclude that QFT of today is not well suited to describe
bound state problem, (see, for example, \cite{Itz}).

The analysis of a bound state is simplest when the constituent
particles can be considered to be nonrelativistic, i.e.,
when they travel at speeds considerably less than $c$. The
physical evident criterion to tell that a bound state is
nonrelativistic is for the binding energy to be small compared
to the rest energies of the constituents. The theoretical
criterion is  that the coupling constants should to be weak and
masses of intermediate particles (photon in QED, mesons in
nuclear physics, gluons in QCD) should be small in comparison
with masses of constituents.

The best example is hydrogen-like systems, which can be
considered nonrelativistic, and the experiments with great
accuracy were required to develop the methods to
calculate next relativistic corrections (see, for example,
\cite{Itz,Sapir,Grein}).

The situation in nuclear and particle physics is completely
different. First, the coupling constants are not small any
more. Second, in nuclear physics, although the binding energy
is relatively small in comparison with nucleon masses,
the masses of intermediate mesons realizing the strong nuclear
interaction are not small. In particular, the most adequate
description of the deuteron can be done by the Bethe-Salpeter
equation where the contribution of all light mesons should be
taken into account (see, for example, \cite{Kapt}).

In particle physics, only hadrons made out of heavy quarks
can be considered by the nonrelativistic potential methods
although the decays of heavy hadrons into light ones require
relativistic methods to describe these transformations.
The most familiar light-quark states are intrinsically
relativistic, so that they require pure
relativistic methods. Besides, they are constituted at
distances where the confinement phenomenon should be
taken into account. In addition, one may ask whether the
free Dirac equation applicable to describe the light quarks
in this region? Therefore, using the nonrelativistic
Schr\"odinger equation to describe the light-quark systems
by fitting parameters of potentials (see, for example,
\cite{Luch,Griff}) can be considered heroic attempt to
understand light meson physics by unsuitable methods in
a very rough theoretical approximation.

In relativistic quantum field theory, bound states are
identified by the occurrence of poles of corresponding
amplitudes or Green functions with appropriate quantum
numbers. These poles have a nonperturbative character, so that
they can arise as a result of a nonperturbative rearrangement
of series over a coupling constant. The investigation of
nonperturbative properties was done by establishing integral
equations among amplitudes and Green functions, using the
specific structure of a Lagrangian. One should say that these
equations, having absolutely general form, in reality can be
used when the kernels contain contributions of the
lowest Feynman diagrams only. It implies that in some sense
the coupling constant should be small enough. The
Bethe-Salpeter equation is the most important integral
equation of this type and it is widely used, especially for
calculation of relativistic corrections in hydrogen-like
systems and deuteron physics (see, for example,
\cite{Itz,Grein,Wick,Nak}).

Now we would like to pay attention to the so-called $Z_2=0$
approach (see, for example, \cite{Hay,Efiv}) which is not
used so widely, although the bosonization  of QCD introducing
bilocal boson-type fields was developed in \cite{Rob}. The
objection and prejudice against this method are based on the
persuasion that in local quantum field theory we should
have local interaction only. For example, the pion as a
quark-antiquark bound state is represented by the local term
$\pi(\bar{q}\gamma_5 q)$. As a result the renormalization
constant $Z_2$ for the pion contains the ultraviolet divergence
and therefore the condition $Z_2=0$ makes no practical sense.
The answer is that the local vertex {\it does not lead} to any
bound state. The vertex connected with the formation of a bound
state {\it should be nonlocal}. Moreover, this nonlocal vertex
in the nonrelativistic limit is directly connected with the
nonrelativistic wave function of this bound state. The question
arises how to find this nonlocal vertex. In the approach
\cite{Rob} this vertex is a solution of the Bethe-Salpeter
equation, which can only be solved by difficult numerical
computations.

Recently, in papers \cite{Efned} we began to develop {\it the
Model of Induced Quark Currents} based on the assumpsion that
(anti-)self-dual homogeneous gluon field realizes the physical
QCD vacuum. To calculate the mass spectrum of bosons as
quark-antiquark bound states, the method of so-called {\it
Bosonization of Nonlocal Currents} (BNC) was used. Numerical
results for the boson spectrum turn out to be in good
agreement with the experimental ones. This method is quite
close to the $Z_2=0$ method and to the bosonization of QCD
introduced by \cite{Rob}. A similar approach was developed for
fermionic many-body problems in statistical physics in
\cite{Kop}. The idea of BNC consists of two point. First, we
write the Green function containing a bound state in the
functional integral representation in which the Gaussian
measure being the lowest order of the bosonization of bilocal
currents contains the Bethe-Salpeter kernel written in
symmetric hermitian form. Second, we have guessed the analytical
form of the orthonormal system of functions very close to the
eigenfunctions of the Bethe-Salpeter equation.

Now we would like to attract attention to the following quite
important point which is not usually stressed clearly enough.
Because exact solutions for any realistic quantum field
model are not known one or another approximation should
be used. In any case, we should have possibilities, at least
in principle, to evaluate the theoretical accuracy of a chosen
approximation. The most simple, although maybe not the most
accurate, way is to require the effective coupling constant to
be noticeably smaller than unity. The perturbation series in
the coupling constant is the most popular method of
calculations, especially in quantum electrodynamics or the
theory of weak interactions. However, if a perturbation series
is rearranged by a subsummation, the problem of the effective
coupling constant becomes quite topical. Thus, if the theory is
reformulated somehow, the effective coupling constant should
be defined for each particular case and should be smaller
than one. Only in this case we can trust our computations.

The aim of this paper is to formulate the BNC-method
and to clarify the connection between the principal
approaches used in atomic, nuclear and particle physics of
today to calculate bound states of interacting particles,
namely, the nonrelativistic potential Schr\"odinger equation
(S-method), the relativistic Bethe-Salpeter equation in the
one-boson-exchange approximation (the BS-method) and our
BNC-method in the one-loop approximation. We answer this
question by considering a relatively simple quantum field
model. The results will provide a deeper understanding
of approximations used in the well-known standard approaches.

An example of the above-mentioned simple quantum field model
is the Yukawa interaction of charged scalar bosons described
by the field $\Phi$ and neutral bosons described by the field
$\phi$. The Lagrangian density is
\begin{eqnarray}
\label{lagr}
L(x)&=&\Phi^+(\Box-M^2)\Phi+{1\over2}\phi(\Box-m^2)
\phi+ g\Phi^+\Phi\phi.
\end{eqnarray}
This model is frequently used as the simplest pattern of QFT
in many discussions, although this system is not stable from a
strict point of view because the Hamiltonian is not bounded
>from below. This model has been investigated by various methods
(see, for example, \cite{Itz,Wick,Bog} and the recent
paper \cite{Dar} and references therein).

In this model it is possible to retrace all details of bound
states arising in quantum field theory. Generalization to the
case of the Dirac field presents no difficulties of principle and
leads to technical problems connected with the algebra of
$\gamma$-matrices only. This model is superrenormalizable so
that the renormalization procedure has the simplest form. The
main aim of this paper is to understand the general mechanism
of bound states arising in this quantum field model and
to clarify conditions on parameters of this quantum
field model dictating a definite method mentioned above to be
applied.

The model contains three dimensionless parameters:
\begin{eqnarray}
\label{param}
&& \lambda={g^2\over16\pi M^2},~~~~\xi={m\over M},
~~~~~b=\left({M_b\over2M}\right)^2=\left(1-{\Delta
M\over2M}\right)^2.
\end{eqnarray}
where $M_b=2M-\Delta M$ is the mass of a supposed bound state,
$\Delta M$
is the mass excess or the binding energy.

Three parameters, $\lambda$, $\xi$ and $b$, are not independent.
The standard formulation of the problem is to find $b$ if
$\xi$ and $\lambda$ are given. We formulate the problem in
another way: what is the region of changing $b$ for a fixed
$\xi$, if the effective coupling constant (which can differ
>from $\lambda$) is smaller than unity?

The parameter $\xi$ is supposed to be smaller than 1. Our aim
is to find the condition under which the mass $M_b$ of a bound
state lies in the interval
\begin{eqnarray*}
&& 0<M_b<2M~~~~~~~{\rm
or}~~~~~~~0<b=\left({M_b\over2M}\right)^2<1,
\end{eqnarray*}
i.e. this bound state should be stable.

It turns out that all approaches mentioned above require
the effective coupling constant to be small enough.
The value of the parameter $\xi={m\over M}$ plays
the crucial role separating nonrelativistic and relativistic
approaches. Namely
\begin{itemize}
\item the {\it Potential nonrelativistic picture} takes place
for
\begin{eqnarray}
\label{potpic}
&& \xi={m\over M}\ll1,~~~~~~~\lambda={g^2\over16\pi M^2}\ll1,
\end{eqnarray}
and, therefore,
$$ 1-b=\left[1-\left({M_b\over2M}\right)^2\right]\ll 1,
~~~~~~{\rm or}~~~~~~\Delta M\ll 2M, $$
i.e., the binding energy is very small;
\item the {\it Bethe-Salpeter approach and Bosonization of
Nonlocal Currents} take place for
\begin{eqnarray}
\label{BNCpic}
&& \xi={m\over M}\leq1,~~~~~~~~~\lambda={g^2\over16\pi M^2}\leq1,
\end{eqnarray}
and the binding energy can be up to $\Delta M\leq(.2\div.3)\cdot
2M$.
\end{itemize}

The result of this paper is that all methods under consideration
can be used in the weak coupling regime only, but the
nonrelativistic potential picture takes place provided the mass
of an exchange particle $m$ to be very small in comparison with
that of the constituent particle $M$. The BS-approach and the
Bosonization of Nonlocal Currents can be used for any relations
between the masses of exchange and constituent particles
although the latter method has a somewhat wider range of
applicability.

From my point of view, it is just the BNC-formalism to be the
most attractive method to study bound states in QFT. It
provides to represent results in an analytic form and
to evaluate the theoretical accuracy of approximations.

This work has been supported by the Russian Foundation for
Fundamental Research N 96-02-17435a.

\vspace{.5cm}

\section{The formulation of the problem}

We consider the interaction of scalar
charged scalar fields $\Phi(x)$ and a neutral scalar field
$\phi(x)$. All the consideration is given in the Euclidean
metrics. The total Lagrangian is done by (\ref{lagr}).
The propagator of a scalar particle with the mass $M$ is
\begin{eqnarray}
\label{prop}
&& D_M(x-y)=\int{dp\over(2\pi)^4}\cdot{e^{ip(x-y)}\over M^2+p^2}.
\end{eqnarray}

The object of our interest is the four-point Green function
\begin{eqnarray}
\label{green4}
&& G(x_1,x_2,y_1,y_2)\\
&& =\int\int D\Phi D\Phi^+D\phi\cdot
\Phi^+(x_1)\Phi(x_2)\Phi^+(y_1)\Phi(y_2)\cdot
e^{S[\Phi^+,\Phi,\phi]},
\nonumber\\
&& ~~~~~~~~~~S[\Phi^+,\Phi,\phi]=\int dx L(x),\nonumber
\end{eqnarray}
where an appropriate normalization should be introduced.

The four-point Green function (\ref{green4}) contains all
information about possible bound states in the channels
$\Phi\Phi$ and $\Phi^+\Phi$. The particles $\Phi$ can be called
constituent particles. We take an interest in bound states in
the channel $\Phi^+\Phi$. The quantum numbers $Q$  can be
fixed by an appropriate vertex
\begin{eqnarray}
&& \Phi^+(x_1)\Phi(x_2)\to J_Q(x)=(\Phi^+ V_Q\Phi)_x=
\Phi^+(x)V_Q(\stackrel{\leftrightarrow}{p}_x)\Phi(x),\nonumber\\
&& ~~~~~~~~~~~~~~~~~\stackrel{\leftrightarrow}{p}_x=
{1\over i}\left[\stackrel{\leftarrow}{\partial}_x-
\stackrel{\rightarrow}{\partial}_x\right].\nonumber
\end{eqnarray}
The nonlocal vertex $V_Q(\stackrel{\leftrightarrow}{p}_x)$
defines the quantum numbers $Q$ of the current
$J_Q=(\Phi^+ V_Q\Phi)$. This vertex can be represented
like
\begin{eqnarray}
\label{vert}
&& V_Q(\stackrel{\leftrightarrow}{p}_x)=
\int du~\tilde{V}_Q(u)e^{iu\stackrel{\leftrightarrow}{p}_x},
\end{eqnarray}
The current $J_Q=(\Phi^+ V_Q\Phi)$ can be written as
\begin{eqnarray}
\label{curr}
J_Q(x)&=&(\Phi^+V_Q\Phi)_x=
(\Phi^+(x)V_Q(\stackrel{\leftrightarrow}{p}_x)\Phi(x))\\
&=&\int du~\Phi^+(x+u)\tilde{V}_Q(u)\Phi(x-u).\nonumber
\end{eqnarray}

The Green function with quantum numbers $Q$ is defined as
\begin{eqnarray}
\label{greenQ}
&& G_Q(x-y)=\int D\Phi D\Phi^+\int D\phi~J_Q(x)J_Q(y)~
e^{S[\Phi^+,\Phi,\phi]}.
\end{eqnarray}
If in this channel a stable bound state with the mass
$M_Q=M_b<2M$ does
exist, the Green function $G_Q(x)$ has the following asymptotic
behaviour
$$ G_Q(x)\sim e^{-M_Q\vert x\vert}~~~~~~{\rm for}~~~~~~
\vert x\vert\to\infty,$$
so that the mass of the state $J_Q=(\Phi^+V_Q\Phi)$ can be found
as
\begin{eqnarray}
\label{massbnd}
&& M_b=M_Q=-\lim_{x\to\infty}{1\over\vert x\vert}\ln G_Q(x).
\end{eqnarray}
The problem is to calculate the functional integral in
representation (\ref{greenQ}) and find the mass $M_b$
according to (\ref{massbnd}).

If we consider perturbation expansion over the coupling constant
$g$ for the four point Green function $G(x_1,x_2,y_1,y_2)$, we
will get a series of the Feynman diagrams describing an
interaction of two particles $\Phi$. This series can
be written in the form of the Bethe-Salpeter equation (see, for
example, \cite{Itz,Nak,Grein}). Bound states of two
particles $\Phi$ in a channel $J_Q(x)=(\Phi^+V_Q\Phi)_x$ can be
found as solutions of this equation.

We proceed in another way. First, our aim is to obtain for the
Green function (\ref{greenQ}) the functional integral
representation in which the term being responsible for bound
state creation would be written in the explicit form and for
remaining corrections the smallness criterion would be defined.
Second, we want to formulate the method of analytical
calculations of binding energies and evaluate their theoretical
accuracy.

Fortunately, it is possible in the representation (\ref{greenQ})
to do the first integration either over the field $\Phi(x)$
or field $\phi(x)$. Thus, we get two representations which
are the starting points of two approaches: the {\it Potential
picture} and the {\it Bosonization of Nonlocal Currents}.

\vspace{.5cm}
{\bf I. Potential picture}
\vspace{.5cm}

The integration in (\ref{green4}) over the charged scalar
field $\Phi$ gives for the Green function (\ref{greenQ}):
\begin{eqnarray}
\label{greenPot}
G_Q(x-y)&=&G_Q^{(P)}(x-y)+G_Q^{(A)}(x-y).
\end{eqnarray}
Here
\begin{eqnarray*}
&& G_Q^{(P)}(x-y)=\int du\int dv\int D\phi~e^{S_P[\phi]}\\
&& ~~~~~~~\cdot\left\{\tilde{V}_Q(u){\bf D}_M(x+u,y+v\vert\phi)
\tilde{V}_Q(v){\bf D}_M(y-v,x-u\vert\phi)\right\},\\
&& G_Q^{(A)}(x-y)=\int du\int dv\int D\phi~e^{S_P[\phi]}~\\
&& ~~~~~~\cdot\left\{\tilde{V}_Q(u){\bf D}_M(x+u,x-u\vert\phi)
\right\}\cdot\left\{\tilde{V}_Q(v){\bf  D}_M(y+v,y-v\vert\phi)
\right\},
\end{eqnarray*}
where
$$ S_P[\phi]={1\over2}\int dx~\phi(x)(\Box-m^2)\phi(x)-
{\rm tr}\ln[1-g\phi D_M].$$
The Green function ${\bf D}_M(x,y\vert\phi)$ satisfies the
equation
\begin{eqnarray}
\label{propPhi}
[-\Box+M^2-g\phi(x)]{\bf D}_M(x,y\vert\phi)=\delta(x-y)
\end{eqnarray}
with
$$ {\bf D}_M^+(x,y\vert\phi)={\bf D}_M(x,y\vert\phi)=
{\bf D}_M(y,x\vert\phi).$$
Here the functions $G_Q^{(P)}(x)$ and $G_Q^{(A)}(x)$ are
said to be "potential" and "annihilation" Green functions,
respectively. The approach based on the representation
(\ref{greenPot}) will be called the {\it Potential picture}.

\vspace{.5cm}
{\bf II. Bosonization of Nonlocal Currents}
\vspace{.5cm}

The integration in (\ref{greenQ}) over the scalar field $\phi(x)$
gives
\begin{eqnarray}
\label{greenBNC}
G_Q(x-y)&=&\int\int D\Phi D\Phi^+\cdot
e^{S_B[\Phi^+,\Phi]}\cdot(\Phi^+V_Q\Phi)_x(\Phi^+V_Q\Phi)_y,
\end{eqnarray}
where an appropriate normalization is implied and
\begin{eqnarray*}
S_B[\Phi^+,\Phi]&=&-(\Phi^+D_M^{-1}\Phi)+
{g^2\over2}(\Phi^+\Phi D_m\Phi^+\Phi),\\
(\Phi^+D_M^{-1}\Phi)&=&\int dx~\Phi^+(x)(-\Box+M^2)\Phi(x),\\
{g^2\over2}(\Phi^+\Phi D_m\Phi^+\Phi)&=&
\int dx\int dy~\Phi^+(x)\Phi(x)D_m(x-y)\Phi^+(y)\Phi(y).
\end{eqnarray*}
The approach based on the representation (\ref{greenBNC}) will
be called {\it the Bosonization of Nonlocal Currents}.

\section{The Potential picture.}

The starting point of the Potential picture is the representation
(\ref{greenPot}).

\subsection{The Green function ${\bf D}_M(x,y|\phi)$}

First, the charged loops should be neglected, so
that the Green functions $G_P(x-y)$ and $G_Q(x-y)$ are
represented by
\begin{eqnarray}
\label{grPA}
G^{(P)}(x-y) &=& \int d\sigma_{uv}[\phi]~
{\bf D}_M(x+u,y+v\vert\phi)\cdot{\bf
D}_M(x-u,y-v\vert\phi),\\
G^{(A)}(x-y) &=& \int d\sigma_{uv}[\phi]~
{\bf D}_M(x+u,x-u\vert\phi)\cdot{\bf D}_M(y+v,y-v\vert\phi)
\nonumber\\
d\sigma_{uv}[\phi] &=& du\tilde{V}(u)~dv\tilde{V}(v)~D\phi~
e^{-{1\over2}(\phi(x)D^{-1}_m\phi)}.
\nonumber
\end{eqnarray}
We would like to stress now that the neglect of loops of
scalar particles $\Phi$ presupposes the dimensionless coupling
constant $\lambda$ to be small enough.

The solution of (\ref{propPhi}) can be represented by
the functional integral (see, for example, \cite{Diney}):
\begin{eqnarray}
\label{prpPhi}
&& {\bf
D}_M(x,y|\phi)={1\over-\Box+M^2-g\phi(x)}\cdot\delta(x-y)\\
&&=\int\limits_0^\infty {d\alpha\over8\pi^2\alpha^2}e^{-
{\alpha\over2}M^2} \int D\xi\exp\left\{-\int\limits_0^\alpha
d\tau{\dot{\xi}^2(\tau)\over2}-{g\over2}\int\limits_0^\alpha
d\tau\phi\left(\xi(\tau)\right)\right\}, \nonumber
\end{eqnarray}
with the boundary conditions $\xi(0)=y,~~\xi(\alpha)=x$ and the
normalization
$$\int D\xi\exp\left\{-\int\limits_0^\alpha
d\tau{\dot{\xi}^2(\tau)\over2}
\right\}=e^{-{(x-y)^2\over2\alpha}}.$$

\subsection{The Green function $G^{(P)}(x)$}

The function $G^{(P)}(x)$ after integration over $\phi$ has the
form
$$ G_P(x)=\left({M\over8\pi^2x}\right)^2
\int d\Sigma_1 d\Sigma_2~e^{W_{11}+2W_{12}+W_{22}},$$
$$ W_{ij}={g^2\over8}\int\limits_{0}^{\alpha_i}d\tau_1
\int\limits_{0}^{\alpha_j}d\tau_2~
D_m\left(\xi_i(\tau_1)-\xi_j(\tau_2)\right),$$
$$\int d\Sigma_j\{\ast\}=
\int du_j\tilde{V}(u_j)\int\limits_0^\infty d\alpha_j~
e^{-{\alpha_j\over2}M^2} \int D\xi_j\exp\left\{-
\int\limits_0^{\alpha_j}
d\tau{\dot{\xi}_j^2(\tau)\over2}\right\}\{\ast\},$$
$$(j=1,2;~~~u_1=u,~~u_2=v),$$
$$ \xi_1(0)=v,~~~~~~\xi_1(\alpha_1)=x+u,~~~~~~~~~~
\xi_2(0)=-v,~~~~~~\xi_2(\alpha_2)=x-u.$$
where for simplicity we put $y=0$.

Our task is to get the asymptotic behaviour of the functions
$G^{(P)}(x)$ for asymptotically large
$x=\sqrt{x^2}\rightarrow\infty$.
To this end, let us introduce
the following variables:
$$ \alpha_j={x\over Ms_j},~~~~~~\tau_j={\beta_j\over Ms_j},$$
$$ \xi_1(\beta)=n\beta+\eta_1(\beta),~~~~~~~
\xi_2(\beta)=n\beta+\eta_2(\beta),~~~~~n_{\mu}=
{x_{\mu}\over x}.$$
Then, one can obtain
\begin{eqnarray}
\label{grP}
G^{(P)}(x)&=&\left({M\over8\pi^2x}\right)^2\int du~\tilde{V}(u)
\int dv~\tilde{V}(v)\\
&\cdot&\int\limits_{0}^{\infty}\int\limits_{0}^{\infty}ds_1ds_2~
e^{-{xM\over2}\left({1\over s_1}+s_1+{1\over s_2}+s_2\right)}
\cdot J^{(P)}(s_1,s_2;x),\nonumber
\end{eqnarray}
where
\begin{eqnarray}
\label{JP}
&& J^{(P)}(s_1,s_2;x)\\
&& =\int\int D\eta_1 D\eta_2\exp\left\{-\int\limits_{0}^{x}d\beta
\left[{Ms_1\dot{\eta}_1^2(\beta)\over2}+
{Ms_2\dot{\eta}_2^2(\beta)\over2}\right]+W_x[\eta_1,\eta_2]\right
\}, \nonumber\\
&&  \eta_1(0)=v,~~~~\eta_1(x)=u,~~~~~~~~~~\eta_2(0)=-v,
~~~\eta_2(x)=-u.\nonumber
\end{eqnarray}
The "potential interaction" is described by the two-point
nonlocal
functional
\begin{eqnarray}
\label{Wij}
&& W_x[\eta_1,\eta_2]=W_{11}+2W_{12}+W_{22},\\
&& W_{ij}={g^2\over8M^2s_is_j}
\int\limits_{0}^{x}\int\limits_{0}^{x}d\beta_1 d\beta_2~
D_m\left(n(\beta_1-\beta_2)+\eta_i(\beta_1)-
\eta_j(\beta_2)\right).\nonumber
\end{eqnarray}
The functional integral for
$J_{x}(s_1,s_2;x)$ looks like the Feynman path integral in the
nonrelativistic statistic quantum mechanics for the four
dimensional motion of particles $\eta_1(\beta)$ and
$\eta_2(\beta)$ with "masses" $Ms_1$ and $Ms_2$ where $\beta$
plays the role of the imaginary time or temperature. The
interaction of these particles is defined by the nonlocal
functional $W_{11}+2W_{12}+W_{22}$ which contains potential
$W_{12}$ and nonpotential $W_{11}+W_{22}$ interactions. It
should be noted that the structure of the functional integral
(\ref{JP}) reminds the polaron problem (see \cite{Feyn}).

The asymptotic form of the function $J^{(P)}(s_1,s_2;x)$ looks
like
\begin{eqnarray}
\label{asJP}
&&J^{(P)}(s_1,s_2;x)\sim\exp\{-x E(s_1,s_2)\},
\end{eqnarray}
where $E(s_1,s_2)$ is the energy of the lowest bound state. The
asymptotic behaviour of the functional $G^{(P)}(x)$ as
$x\rightarrow\infty$ is determined by the saddle point of the
integrals over $s_1$ and $s_2$ in the representation
(\ref{grP}). Substituting expression (\ref{asJP}) into
(\ref{grP}), one can get
\begin{eqnarray}
\label{mnbdP}
M_b&=&-\lim_{x\to\infty}{1\over\vert x\vert}\ln G^{(P)}(x),\\
&=&\min_{(s_1, s_2)}\left[{M\over2}\left({1\over s_1}+s_1
+{1\over s_2}+s_2\right)+E(s_1,s_2)\right]\nonumber\\
&=&\min_s\left[M\left({1\over
s}+s\right)+E(s,s)\right].\nonumber
\end{eqnarray}

The main problem is to compute the functional integral
(\ref{JP}). This computation can be done by the variational
methods (see \cite{Feyn}). We plan to calculate this functional
integral applying the Gaussian equivalent representation method
which was sucessfully used sucessfully for the polaron problem
(see \cite{Diney}).

Besides one can see that the representations (\ref{grP}) and
(\ref{JP}) does not really feel the explicit form of the
vertex $\tilde{V}_Q(u)$ although it should extract a bound
state with definite quantum numbers $Q$. It means, in fact,
that in the general case, i.e., for any value of the coupling
constant $g$, the modern analytical methods, applied to the
functional integral (\ref{grP}), allow to calculate with
reasonable accuracy the energy of the lowest bound state only.

\subsection{The Green function $G^{(A)}(x-y)$}

The function $G^{(A)}(x)$ after integration over $\phi$ has the
form
$$ G^{(A)}(x)=\left({M\over8\pi^2x}\right)^2
\int d\Sigma_1 d\Sigma_2 e^{W_{11}+2W_{12}+W_{22}},$$
$$ W_{jj}={g^2\over2}\int\limits_{0}^{\alpha_j}d\tau_1
\int\limits_{0}^{\alpha_j}d\tau_2
D_m\left(\xi_j(\tau_1)-\xi_j(\tau_2)\right),~~~~~(j=1,2),$$ $$
W_{12}={g^2\over2}\int\limits_{0}^{\alpha_1}d\tau_1
\int\limits_{0}^{\alpha_2}d\tau_2
D_m\left(x+\xi_1(\tau_1)-\xi_2(\tau_2)\right).$$
Here $ \xi_1(0)=-u,~~~~~\xi_1(\alpha_1)=u,~~~~~
\xi_2(0)=-v,~~~~~\xi_2(\alpha_2)=v.$

One can see that in the limit $x\to\infty$ $ W_{12}\to 0$ and
$$ G^{(A)}(x)\to\left[{M\over8\pi^2x}
\int d\Sigma_1e^{W_{11}}\right]^2,$$
so that no bound state arises in this case. Thus, the
annihilation
channel does not contain any bound states. In other words,
intermediate pure boson states of particles $\phi$ cannot
produce any bound state.

\subsection{ The Nonrelativistic Limit }

In this section we obtain the nonrelativistic limit
$c\to\infty$ for the loop function $G_P(x)$ in
(\ref{grP}). Our task is to introduce the
parameter $c$ in an explicit form into $G_P(x)$
and then go to the limit $c\to\infty$.
To this end, let us restore the parameter $c$ in
our formulas. We have
\begin{eqnarray}
\label{c}
&& M\to Mc,~~~~ g\to g ,
\end{eqnarray}
$$ x_\mu=(x_4,{\bf x})\to(ct,{\bf x}),~~~~
 x=\sqrt{x^2}=\sqrt{c^2t^2+{\bf x}^2}\to ct,$$
$$ n_\mu={x_\mu\over\sqrt{x^2}}\to(1,{\bf x}/ct)\to (1,0).$$
The propagator $D_m(x)$ becomes
\begin{eqnarray}
\label{Dcinf}
D_m(x)&\to& cD_\kappa(ct,{\bf x})\\
&=&c\int{d^4 k\over{(2\pi)^4}}
\tilde{D}_m(k^2)e^{i(k_4t+{\bf k}{\bf x})}
=\int {d{\bf k}\over{(2\pi)^3}}
\int\limits_{-\infty}^{\infty}{dv\over{2\pi}}
\cdot{e^{i(vt+{\bf k}{\bf x})}\over{\bf k}^2+\kappa^2+
{v^2\over{c^2}}} \nonumber\\
&\to&\int {d{\bf k}\over{(2\pi)^3}}
\int\limits_{-\infty}^{\infty}{dv\over{2\pi}}
\cdot{e^{i(vt+{\bf k}{\bf x})}\over{\bf k}^2+\kappa^2}
\cdot\left[1-{v^2\over c^2}\cdot
{1\over{\bf k}^2+\kappa^2}\right]\nonumber\\
&=&\delta(t)\cdot{1\over4\pi}{e^{-\kappa r}\over r}+
{1\over c^2}\delta''(t)\cdot{1\over8\pi}
{e^{-\kappa r}\over\kappa}+O\left({1\over c^4}\right).
\nonumber
\end{eqnarray}
The parameter ${1\over\kappa}$ defines the radius of the
nonrelativistic Yukawa potential, therefore, we have to keep
$$\kappa=mc={\rm const}$$
to be finite in the limit $c\to\infty$.

Let us come back to the Green function (\ref{grP}).
According to (\ref{c}) it reads
\begin{eqnarray*}
G_P(t)&=&\left({M\over4\pi t}\right)^2
\int\!\!\!\!\int\limits_0^\infty
ds_1ds_2~e^{-t{Mc^2\over 2}\left({1\over s_1}+s_1
+{1\over s_2}+s_2\right)}J_P(s_1,s_2;t)
\end{eqnarray*}
In the functional integral (\ref{JP}) let us introduce
the new varibles $\beta_j=c\tau_j$. It is convinient
to represent the four-vectors $\eta_j$ in the one and
three component form
$$ \eta_j(\beta)=(a_j(\tau),{\bf b}_j(\tau)),~~~~~(j=1,2)$$
and
$$ a_1(\tau)=\Phi(\tau)+{\phi(\tau)\over2},~~~~~~~~
a_2(\tau)=\Phi(\tau)-{\phi(\tau)\over2},$$
$${\bf b}_1(\tau)={\bf R}(\tau)+{{\bf r}(\tau)\over2},~~~~~~~~
{\bf b}_2(\tau)={\bf R}(\tau)-{{\bf r}(\tau)\over2}.$$
with the boundary conditions
$$ {\bf R}(0)={\bf R}(t)=0,~~~~~~~
{\bf r}(0)=2{\bf u},~~~~{\bf r}(t)=2{\bf v} $$
In the limit $c\to\infty$ the integral for $G_P(t)$ over
$s_1$ and $s_2$ can be calculated by the saddle-point method.
The saddle points are $s_1=s_2=1$ and we get
\begin{eqnarray}
\label{grP2}
G_P(t)&\approx&{M\over8\pi t^3c^2}e^{-2Mc^2t}\cdot J_P(t)
\end{eqnarray}
where the functional integral looks
\begin{eqnarray*}
J_P(t)&=&\int\!\!\int Da_1Da_2
\int\!\!\int D{\bf b}_1D{\bf b}_2~
e^{-{M\over2}\int\limits_{0}^{t}d\tau
\left[\dot{a}_1^2+\dot{a}_2^2+
\dot{{\bf b}}_1^2+\dot{{\bf b}}_2^2\right]+
W_{11}+2W_{12}+W_{22}}\\
&=&\int\!\!\int D\Phi D\phi\int\!\!\int D{\bf R}D{\bf r}~
e^{-M\int\limits_{0}^{t}d\tau
\left[\dot{\Phi}^2+{1\over4}\dot{\phi}^2+
\dot{{\bf R}}^2+{1\over4}\dot{{\bf r}}^2\right]
+W_{11}+2W_{12}+W_{22}}.
\end{eqnarray*}

The functionals $W_{11}$ and $W_{22}$ describe the
mass renormalization in the nonrelativistic limit and
we omit them. According to (\ref{c}) and (\ref{Dcinf})
the functional $W_{12}$ (\ref{Wij}), where the argument
of $D$-function is
$$ n(\beta_1-\beta_2)+\eta_1(\beta)-\eta_2(\beta)=
(c(\tau_1-\tau_2)+a_1(\tau_1)-a_2(\tau_2),
{\bf b}_1(\tau_1)-{\bf b}_2(\tau_2)),$$
acquires the form in the limit $c\to\infty$
\begin{eqnarray*}
2W_{12}&=&{g^2\over16\pi M^2}\int\!\!\!\!\int\limits_0^t
d\tau_1 d\tau_2\left\{\left[\delta(\tau_1-\tau_2)+{1\over c}
\delta'(\tau_1-\tau_2)(a_1(\tau_1)-a_2(\tau_2))\right]
\cdot{e^{-\kappa r}\over r}\right.\\
&+&\left.{1\over 2c^2}\delta''(\tau_1-\tau_2)\cdot
{e^{-\kappa r}\over\kappa}+...\right\}\\
&=&\lambda\int\limits_0^td\tau~
\left\{\left[1+{1\over 8c^2}\left(\dot{{\bf r}}^2-
{(\dot{{\bf r}}{\bf r})^2\over r^2}(1+\kappa r)\right)
\right]\cdot{e^{-\kappa r}\over r}
-{1\over c}\dot{\Phi}{e^{-\kappa r}\over r}\right.\\
&+&\left. O\left({1\over c}\phi\dot{{\bf R}},{1\over c^2}
\dot{{\bf R}}^2\right)\right\}
\end{eqnarray*}
where $r=r(\tau)=\vert{\bf b}_1(\tau)-{\bf b}_2(\tau)\vert$.

After integration over $\Phi$, $\phi$ and ${\bf R}$
in the functional integral for $J_P$ one can get
\begin{eqnarray}
\label{potY}
J_P(t)&=&\int\limits_{{\bf r}(0)=2{\bf u}}^{{\bf r}(t)=2{\bf v}}
D{\bf r}~\cdot\exp\left\{-\int\limits_0^t d\tau~\left[{M_c\over2}
\dot{{\bf r}}^2(\tau)+U({\bf r}(\tau))\right]\right\},
\\
U({\bf r}))&=&-\lambda{e^{-\kappa r}\over r}
-{\lambda\over2M_c^2c^2}p_i\left(\left[\delta_{ij}-
{r_ir_j\over r^2}(1+\kappa r)\right]\cdot
{e^{-\kappa r}\over r}\right)p_j
-{\lambda^2\over8M_cc^2}{e^{-2\kappa r}\over r^2}\nonumber\\
\end{eqnarray}
where $M_c={M\over2}$ and $p_j=M_c\dot{r}_j$.

Here we keep the main terms contributing to the potential and
neglect terms describing and the nonlocal interaction
and terms of the order $O(1/c^4)$.

One can see that this representation for $J_P(t)$ coincides with
the Feynman path integral in quantum machanics for the Green
function
$K({\bf v},t;{\bf u},0)$
\begin{eqnarray}
\label{Fn}
K({\bf v},t;{\bf u},0)&=&
\int\limits_{{\bf r}(0)=2{\bf u}}^{{\bf r}(t)=2{\bf v}}D{\bf r}
\exp\left\{-\int\limits_0^td\tau\left[{M_{cm}\over2}
\dot{{\bf r}}^2(\tau)+U({\bf r}(\tau))\right]\right\},
\\
&=&\sum\limits_Q\psi_Q(\vec{v})e^{-tE_Q}\psi_Q(\vec{u})
\nonumber
\end{eqnarray}
Here $\psi_Q({\bf r})$ and $E_Q$ are eigenfunctions and
eigenvalues for the quantun number $Q$ connected with the
space ${\bf R}^3$ of the Schr\"{o}dinger equation
\begin{eqnarray}
\label{Seq}
\left[{p^2\over 2M_{cm}}+U({\bf r})\right]
\psi_Q(\vec{r})=E_Q\psi_Q({\bf r})
\end{eqnarray}
where $U({\bf r})<0$ is the attractive potential.

As a result, the Green function $G_P(t)$ for $t\to\infty$
behaves like
\begin{eqnarray}
&& G_P(t)=\sum\limits_Qe^{-tE_Q}
\left[\int d{\bf u}~\tilde{V}({\bf u})\psi_Q({\bf u})\right]^2.
\nonumber
\end{eqnarray}
If we choose
$$\tilde{V}({\bf u})=\int du_4~\tilde{V}(u_4,{\bf u})=
\psi_{Q_0}({\bf u}),$$
then
$$\int d{\bf u}~\psi_Q({\bf u})\psi_{Q_0}({\bf u})=
\delta_{QQ_0},$$
and, finally, for large $t$ we have
\begin{eqnarray*}
&& G_P(t) \rightarrow e^{-tE_{Q_0}},
\end{eqnarray*}
where $E_{Q_0}$ is the energy of the bound state of two
nonrelativistic particles in the quantum state $Q_0$ arising
due to the potential $U({\bf r})$. The mass of the bound
state in the nonrelativistic approach is
\begin{eqnarray}
\label{mSeq}
M_{Q_0}&=&{1\over c^2}\left[Mc^2+E_{Q_0}+
O\left({1\over c}\right)\right]
=2M+{E_{Q_0}\over c^2}+O\left({1\over c^3}\right).
\end{eqnarray}

Thus, in the nonrelativitic limit the relativistic vertex
$\tilde{V}(u)$ is connected with the nonrelativistic
eigenfunction
\begin{eqnarray}
\label{vrtvf}
&&\tilde{V}(u)\to\int du_4~\tilde{V}(u_4,{\bf u})=
\tilde{V}({\bf u})=\psi_Q({\bf u})
\end{eqnarray}
and the bound state mass of two scalar particles is a sum
of their masses plus the binding energy $E_Q$ determined by
the nonrelativistic potential interaction.

\subsection{ The nonrelativistic Yukawa potential and
relativistic corrections.}

Thus in the nonrelativistic limit, we have the Schr\"{o}dinger
equation with relativistic corrections:
\begin{eqnarray}
\label{Ham}
&& H=H_Y+{1\over c^2}(U_p+U_s)+O\left({1\over c^4}\right)
\end{eqnarray}
where $H_Y$ is the Yukawa potential
\begin{eqnarray*}
&& H_Y={p^2\over 2M_{cm}}-\lambda{e^{-mr}\over
r},~~~~~M_{cm}={M\over2},~~~~~m=\kappa,
\end{eqnarray*}
and $U_p$ and $U_s$ are the first and second lowest
relativistic corrections in (\ref{potY}).

Our problem is to find restrictions on the parameters $\lambda$
and $\xi={m\over M}$ for which the ralativistic corrections
$U_p$ and $U_s$ can be neglected. We proceed in the simplest way.
We use the variation function
$$ \Psi(r)=e^{-{1\over2}msr},$$
where $s$ is the variational parameter, for the ground state.
This approach gives us quite a good
qualitative and even semi-quantitative estimation of the
background  energy. It is sufficient for our aim. The
calculations give
\begin{eqnarray*}
E_Y&\approx&\min\limits_a{\langle\Psi\vert
H_Y\vert\Psi\rangle\over\langle\Psi\vert\Psi\rangle}
={m^2\over4M}\cdot\min\limits_s\left\{
s^2-G\cdot{s^3\over(1+s)^2}\right\}\\
&=&-{m^2\over8M_c}\cdot{s^2(s-1)\over3+s},
\end{eqnarray*}
where the parameter $s$ is defined by the equation
\begin{eqnarray*}
G&=&4\lambda\cdot{M_c\over m}={2(1+s)^3\over s(3+s)}.
\end{eqnarray*}
The bound state can exist if $s>1$, i.e.
$$ G>4,~~~~~~~{\rm or}~~~~~~~\lambda>{m\over M_c}$$
The contribution of relativistic corrections can be
evaluated as follows:
\begin{eqnarray*}
\delta_p&=&{\langle\Psi\vert U_p\vert\Psi\rangle\over
\langle\Psi\vert U_Y\vert\Psi\rangle}=
{1\over4}\cdot\left({m\over M}\right)^2\cdot{s^2\over1+s},\\
\delta_s&=&{\langle\Psi\vert U_s\vert\Psi\rangle\over
\langle\Psi\vert U_Y\vert\Psi\rangle}={\lambda\over8}\cdot
{m\over M}\cdot{(1+s)^2\over2+s},
\end{eqnarray*}
Thus the nonrelativistic picture takes place if
$$ \vert E_Y\vert\ll2M,~~~~~~\delta_p\ll1,~~~~~~\delta_s\ll1$$
It is easy to see that these inequalities take place if
\begin{eqnarray}
\label{Prestr}
&& \xi={m\over M}\ll1,~~~~~~~~~\lambda\ll1.
\end{eqnarray}
It is valid for $s\sim2\div3$ and $a\sim (1\div2)\cdot m$.

\subsection{Relativistic incompleteness
of quantum mechanics of two particles.}

Here we would like to pay attention to the Schr\"{o}dinger
equation which describes two nonrelativistic particles
\begin{eqnarray}
&& H={p_1^2\over2m_1}+{p_2^2\over2m_2}-U(\vec{r}_1-\vec{r}_2),
\nonumber
\end{eqnarray}
where the potential is attractive.

Let us pass to the center-of-mass system in a standard way
$$ \vec{R}={m_1\vec{r}_1+m_2\vec{r}_2\over m_1+m_2},~~~~~~~
\vec{r}=\vec{r}_1-\vec{r}_2. $$
The Hamiltonian takes the form
\begin{eqnarray}
&& H={p^2\over2M}+{p_r^2\over2\mu}-U(r)\nonumber
\end{eqnarray}
$$ M=m_1+m_2,~~~~~~~\mu={m_1m_2\over m_1+m_2} $$
The solution of the Schr\"{o}dinger equation $H\Psi=E\Psi$
can be written as
$$ \Psi(\vec{R},\vec{r})=e^{i\vec{p}\vec{R}}\psi(r)$$
where $\vec{p}$ is the momentum of the total system;
$\psi(r)$ is an eigenfunction of the equation
\begin{eqnarray*}
&& \left[{\vec{p}_r^2\over2\mu}-U(r)\right]\psi(r)
=-\varepsilon\psi(r),
\end{eqnarray*}
and $-\varepsilon~(\varepsilon>0)$ is an eigenvalue of a bound
state. Then, the eigenvalue or the energy of the state
$\Psi(\vec{R},\vec{r})$ for $\vec{p}\neq 0$ is
\begin{eqnarray*}
&& E={p^2\over2M}-\varepsilon.
\end{eqnarray*}
From the physical point of view this energy has no reasonable
meaning. Indeed, we should get
\begin{eqnarray*}
&& E={p^2\over2M}-\varepsilon\to{p^2\over2M_{phys}},\\
&& M_{phys}=m_1+m_2-\Delta,~~~~~~~\Delta={\varepsilon\over c^2}
\end{eqnarray*}
i.e., the interaction between two particles should give the mass
excess.

On the other hand, the latter formula can be obtained from the
relativistic energy in the nonrelativistic limit
\begin{eqnarray*}
E&=&\sqrt{M_{phys}^2c^4+p^2c^2}=
M_{phys}c^2+{p^2\over2M_{phys}}+O(p^2)\\
&=&(m_1+m_2-\Delta M)c^2+{p^2\over(m_1+m_2-\Delta M)}+O(p^2)\\
&=&(m_1+m_2)c^2+{p^2\over2(m_1+m_2)}-\Delta M c^2+
O\left({\Delta M\over M}\right)+O(p^2)
\end{eqnarray*}
and the mass of the bound state equals
\begin{eqnarray*}
&& M_{phys}=M-{\varepsilon\over c^2}+O\left({1\over c^4}\right).
\end{eqnarray*}
Thus, the nonrelativistic Schr\"{o}dinger equation describing two
nonrelativistic particles can be cosidered as a fragment of an
relativistic equation describing the relativistic interaction of
two particles. The "true" relativistic equation should contain
terms describing both the motion of the center of masses and
their relative motion.

\section{Bosonization of Nonlocal Currents.}

The starting point of {\it Bosonization of charged currents}
is the representation (\ref{greenBNC}).

\subsection{Bilocal currents.}

Let us consider the four-field term (\ref{greenBNC}) and
introduce the bilocal currents:
\begin{eqnarray*}
&&{g^2\over2}(\Phi^+\Phi D_m\Phi^+\Phi)\\
&&={g^2\over2}\int\int dy_1dy_2~\Phi^+(y_1)\Phi(y_1)
D_m(y_1-y_2)\Phi^+(y_2)\Phi(y_2)\\
&&={g^2\over2}\int\int dy_1dy_2~J^+(y_1,y_2)J(y_1,y_2)
={g^2\over2}(J^+J)
\end{eqnarray*}
where
\begin{eqnarray}
\label{blocJ}
J(y_1,y_2)&=&\sqrt{D_m(y_1-y_2)}(\Phi^+(y_1)\Phi(y_2)),\\
&& J^+(y_1,y_2)=J(y_2,y_1).\nonumber
\end{eqnarray}

The next step is to use the Gaussian representation
\begin{eqnarray*}
&& e^{{g^2\over2}(\Phi^+\Phi D_m\Phi^+\Phi)}=e^{{g^2\over2}
(J^+J)}=\int DA~e^{-{1\over2}(A^+A)-g~(A^+J)}.
\end{eqnarray*}
where  $A^+(y_1,y_2)=A(y_2,y_1)$ and
\begin{eqnarray*}
DA&=&\prod_{y_1,y_2}dA(y_1,y_2),\\
(A^+A)&=&\int\int dy_1dy_2~A^+(y_1,y_2)A(y_1,y_2),\\
(A^+J)&=&\int\int dy_1dy_2~A^+(y_1,y_2)J(y_1,y_2)=(AJ^+)\\
&=&\int\int dy_1dy_2~\Phi^+(y_1)A^+(y_1,y_2)
\sqrt{D_m(y_1-y_2)}\Phi(y_2)\\
&=&(\Phi^+(A^+\sqrt{D_m})\Phi).
\end{eqnarray*}

We can always represent the product
$\Phi^+(x_1)\Phi(x_2)\Phi^+(x_3)\Phi(x_4)$
as a combination of currents
\begin{eqnarray*}
g^2\Phi^+(x_1)\Phi(x_2)\Phi^+(x_3)\Phi(x_4)\to
g^2J(x_1,x_2)J(x_3,x_4).
\end{eqnarray*}
We have
\begin{eqnarray*}
&& G(x_1,x_2;x_3,x_4)\\
&=&\int D\Phi D\Phi^+\int DA~g^2J(x_1,x_2)J(x_3,x_4)
\cdot e^{-(\Phi^+D^{-1}_M\Phi)-{1\over2}(A^+A)-g~(A^+J)}.
\end{eqnarray*}
Now we integrate by parts over $A$ and $A^+$ and calculate the
Gaussian integral over $\Phi$ and $\Phi^+$:
\begin{eqnarray}
\label{grnAA}
&& G(x_1,x_2;x_3,x_4)=\int\int DA~A(x_1,x_2)A(x_3,x_4)
~e^{S[A]},\\
&& S[A]=-{1\over2}(A^+A)
-{\rm tr}\ln\left[1+g(A^+\sqrt{D_m})D_M\right].\nonumber
\end{eqnarray}
Matrix operations are defined by formulas
$$\left(A^+\sqrt{D_m}\right)_{x_1,x_2}=A^+(x_1,x_2)
\sqrt{D_m(x_1-x_2)},$$
$$\left[(A^+\sqrt{D_m})D_M\right]_{x_1,x_2}=
\int dy~A^+(x_1,y)\sqrt{D_m(x_1-y)}D_M(y-x_2).$$

We would like to stress that the representation
(\ref{grnAA}) is completely equivalent to the initial
representation (\ref{greenBNC}). The Green function
$G(x_1,x_2;x_3,x_4)$ can be considered as the Green function of
the bilocal field $A(x_1,x_2)$. These fields are described by
the nonlocal action $S[A]$.

\subsection{One-loop representation.}

The next problem is to give the standard particle interpretation
to the action $S[A]$ in (\ref{grnAA}). For this aim this action
should be represented in the form
$$ S[A]\to-{1\over2}(A^+R^{-1}A)+I_{int}[A],~~~~~~~
I_{int}[A]=O(A^3).$$
It means that we have to remove the term linear in $A$ and
extract the quadratic term out of $S[A]$. Let us introduce the
displacement
\begin{eqnarray*}
&& A(y_1,y_2)\to A(y_1,y_2)+A_0(y_1,y_2)\\
\end{eqnarray*}
We get
\begin{eqnarray}
\label{actnA}
&& S[A]=-E_0-{1\over2}(A^+A)-(A^+A_0)
-{\rm tr}\ln\left[1+g\left(A^+\sqrt{D_m}\right){\cal D}\right],
\end{eqnarray}
\begin{eqnarray*}
&& {\cal D}=D_M\cdot{1\over1+g(A_0^+\sqrt{D_m})D_M}
\end{eqnarray*}
where the matrix multiplication is implied.

The constant term $E_0$ is the vacuum energy in the lowest
approximation:
\begin{eqnarray*}
&& E_0=-{1\over2}(A_0^+A_0)+
{\rm tr}\ln\left[1+g\left(A_0^+\sqrt{D_m}\right)D_M\right]
\end{eqnarray*}
and will be omitted in subsequent calculations.

\subsubsection{Linear term.}

The term linear in $A$ should be equal to zero
\begin{eqnarray*}
&& (A^+A_0)+g~{\rm tr}[(A^+\sqrt{D_m}){\cal D}]=0,
\end{eqnarray*}
or
\begin{eqnarray*}
&& A_0(x_1,x_2)+g\sqrt{D_m(x_1-x_2)}{\cal D}(x_1,x_2)=0
\end{eqnarray*}
Introducing the function
$$ A_0(x_1,x_2)=A_0(x_1-x_2)={a(x_1-x_2)\over
g\sqrt{D_m(x_1-x_2)}}$$
we get the equation
\begin{eqnarray*}
&& a(x_1-x_2)=-g^2D_m(x_1-x_2){\cal D}(x_1-x_2),
\end{eqnarray*}
where
\begin{eqnarray*}
&& {\cal D}=D_M\cdot{1\over1+aD_M},~~~~~~~~
\tilde{{\cal D}}(k^2)={1\over M^2+k^2+\tilde{a}(k^2)}.
\end{eqnarray*}
Finally, we arrive at the equation of the Schwinger-Dyson type:
\begin{eqnarray}
\label{SDeq}
&& \tilde{a}(k^2)=-g^2\int{dp\over(2\pi)^4}\cdot
{1\over(m^2+(k-p)^2)(M^2+p^2+\tilde{a}(p^2))}.
\end{eqnarray}

The integral in this equation contains the logarithmic
ultraviolet divergence which can be removed by the
renormalization of the mass $M$. It means that we should put
\begin{eqnarray*}
&& M^2+\tilde{a}(k^2)=M_r^2+\tilde{a}_r(k^2,M_r^2),
\end{eqnarray*}
where $M_r$ is the "physical" mass of the constituent particle
$\Phi$
and
\begin{eqnarray*}
&& \tilde{a}_r(k^2,M_r^2)=\tilde{a}(k^2)-\tilde{a}(-M_r^2).
\end{eqnarray*}
It is convinient to define the dimensionless function
$$w(k^2)={\tilde{a}^{(r)}(k^2,M_r^2)\over M_r^2}$$
which satisfies the equation
\begin{eqnarray}
\label{SDeqw}
w(k^2)&=&{g^2\over M_r^2}\int{dp\over(2\pi)^4}\cdot\left[\left.
{1\over(m^2+(q-p)^2)(M_r^2+p^2+M_r^2w(p^2))}\right\vert_{q^2=-
M_r^2}\right.
\\
&-&\left.{1\over(m^2+(k-p)^2)(M_r^2+p^2+M_r^2w(p^2))}\right].\nonumber
\end{eqnarray}

We have obtained the functional equation of the type
$$ w(k^2)=F[w,k^2].$$
Solution can be found by the fixed point method, i.e.
we choose the initial "point" $w_0(k^2)$ and calculate
$$ w_{n+1}(k^2)=F[w_n,k^2],~~~~~~{\rm for}~~~~~~n=0,1,2,...  .$$
In the limit $n\to\infty$ we get
$$ w_n(k^2)\to w(k^2).$$
The main problem is to choose the zeroth approximation
$w_0(k^2)$. We proceed in the following way. The propagator
$\tilde{{\cal D}}(k^2)$ after the mass renormalization should
behave for $k^2\to M_r^2$ like
\begin{eqnarray}
\label{prp0}
&& \tilde{{\cal D}}(k^2)={1\over M_r^2+k^2+M_r^2w(k^2)}\to
{Z\over M_r^2+k^2},\\
&& Z={1\over1+c},~~~~~~~\left.c=
M_r^2{d\over dk^2}w(k^2)\right\vert_{k^2=-M_r^2}\nonumber
\end{eqnarray}
Let us define the zeroth approximation
$$ \tilde{{\cal D}}_{(0)}(k^2)={Z\over
M_r^2+k^2},~~~~~~~Z={1\over1+c}$$
where the constant $c$ is determined by the equation
\begin{eqnarray*}
&&c=\left.M_r^2{d\over
dk^2}w_{(0)}(k^2)\right\vert_{k^2=-M_r^2}\\
&&=-\left.M_r^2{d\over dk^2}\left\{g^2\int{dp\over(2\pi)^4}\cdot
{1\over(m^2+(k-p)^2)}\cdot{Z\over M_r^2+p^2}\right\}
\right\vert_{k^2=-M_r^2}\\
&&=Z{g^2\over16\pi M_r^2}\cdot{1\over\pi}\int\limits_0^1
{d\alpha~\alpha(1-\alpha)\over \xi^2\alpha+(1-\alpha)^2}
=\lambda Z\cdot a(\xi).
\end{eqnarray*}
Here
\begin{eqnarray}
\label{fna}
&& a(\xi)={1\over\pi}\left[-1-(1-\xi^2)\ln\xi
+{\xi(3-\xi^2)\over\sqrt{4-\xi^2}}\cdot
\arctan\left({\sqrt{4-\xi^2}\over\xi}\right)\right].
\end{eqnarray}
The equation
$$Z={1\over1+c}={1\over1+\lambda Za(\xi)}$$
gives
\begin{eqnarray}
&& Z=Z_{(0)}={2\over1+\sqrt{1+4\lambda a(\xi)}}.
\end{eqnarray}

Thus, one can choose in the zeroth approximation
$$ w_{(0)}(p^2)=\left(1+{k^2\over M_r^2}\right)\left({1\over
Z}-1\right).$$

In the first approximation (n=1) we have
\begin{eqnarray*}
w_{(1)}(k^2)&=& \lambda Z\int\limits_0^1{d\alpha\over\pi}~
\ln\left\{{m^2\alpha+M_r^2(1-\alpha)+\alpha(1-\alpha)k^2\over
m^2\alpha+M_r^2(1-\alpha)^2}\right\}
\end{eqnarray*}
where
\begin{eqnarray*}
c_{(1)}&=&\left.M_r^2{d\over
dk^2}w_{(1)}(k^2)\right\vert_{k^2=-M_r^2}=
\lambda Z_{(0)}a(\xi)=c,\\
&&Z_{(1)}={1\over1+c_{(1)}}=Z_{(0)}.
\end{eqnarray*}

Finally, we have in the zeroth and first approximations
\begin{eqnarray}
\label{prp01}
&& \tilde{{\cal D}}_{(0)}(k^2)={Z\over M_r^2+k^2},\\
&& \tilde{{\cal D}}_{(1)}(k^2)={1\over
M_r^2+k^2+M_r^2w_{(1)}(k^2)}\to
{Z\over M_r^2+k^2}~~~{\rm for}~~~k^2\to M_r^2.\nonumber
\end{eqnarray}
One can check that
$$ \tilde{{\cal D}}_{(0)}(k^2)\approx
\tilde{{\cal D}}_{(1)}(k^2)$$
with 5\% accuracy.

The renormalized coupling constant $g_r$ can be defined by a
standard way:
\begin{eqnarray}
\label{rencc}
&& g_r=gZ,~~~~~~~~\lambda_r=\lambda Z^2
\end{eqnarray}
because
\begin{eqnarray*}
&& g\tilde{{\it D}}(k^2)={g\over M_r^2+k^2+M_r^2w(k^2)}
\to{gZ\over M_r^2+k^2}~~~{\rm for}~~~k^2\to M_r^2.
\end{eqnarray*}

In subsequent numerical calculations we use the zeroth
approximation ${\it D}_{(0)}$ (\ref{prp01}) which gives quite
acceptable qualitative semiquantative estimations.

In this approximation the renormalized coupling constant is
\begin{eqnarray}
\label{renccapp}
&&\lambda_r=\lambda_r(\lambda,\xi)=\lambda Z^2
={4\lambda\over(1+\sqrt{1+4\lambda a(\xi)})^2}
\end{eqnarray}
It is important that the renormalization considerably
diminishes the initial coupling constant $\lambda$. For example,
in the case of the "deuteron", when $\Phi$-particle is the
"proton",
$\phi$-paricle is the "$\pi$-meson" and $\lambda=14.5$
$$\lambda_r=\lambda_r\left(14.5,{m_\pi\over M_p}\right)=1.712.$$
Moreover, the renormalized coupling constant $\lambda_r$
is bounded for any $\lambda$
\begin{eqnarray*}
&&\lambda_r\leq{1\over a(\xi)}\leq {1\over a(1)}=15.01... .
\end{eqnarray*}

\subsubsection{From bilocal to local fields}

After removing the linear term we have
\begin{eqnarray*}
S[A]&=&-{1\over2}(A^+A)
-{\rm tr}\ln_1\left[1+g(A^+\sqrt{D_m}){\cal D}\right],\\
&& \ln_1(1+Q)=\ln(1+Q)-Q.
\end{eqnarray*}

The "trace" of "ln" consists of terms
\begin{eqnarray}
\label{term0}
&&\int\int dx_jdx_{j+1}~{\cal D}(...-x_j)A^+(x_j,x_{j+1})
\sqrt{D_m(x_j-x_{j+1})}{\cal D}(x_{j+1}-...)
\end{eqnarray}
Let us proceed as follows. We introduce new variables
$$ x_j=z_j+{y_j\over2},~~~~~~~~~~x_{j+1}=z_j-{y_j\over2};$$
and the notation
$$ A(x_j,x_{j+1})=W(z_j,y_j),~~~~
A(x_{j+1},x_j)=A^+(x_j,x_{j+1})=W(z_j,-y_j).$$
The term (\ref{term0}) can be written as
\begin{eqnarray}
\label{term1}
&&\int\int dz_jdy_j~{\it D}\left(...-z_j-{y_j\over2}\right)
W(z_j,-y_j)\sqrt{D_m(y_j)}{\it D}\left(z_j-{y_j\over2}...\right)
\\
&&=\int dz_j~{\it D}(...-z_j)
{\cal V}(z_j,\stackrel{\leftrightarrow}{p}_{z_j}){\it D}(z_j-...)
\nonumber
\end{eqnarray}
where
$${\cal V}(z,\stackrel{\leftrightarrow}{p}_{z})=
\int dy~\sqrt{D_m(y)}W(z,-y)
e^{-i{y\over2}\stackrel{\leftrightarrow}{p}_{z}},$$
$$ \stackrel{\leftrightarrow}{p}_{x}=
{1\over i}\left[\stackrel{\leftarrow}{\partial}_{x}-
\stackrel{\rightarrow}{\partial}_{x}\right] $$

Let the system of functions $\{U_Q(y)\}$ with quantum numbers
$Q=(nl\{\mu\})$, where $n$, $l$ and $\{\mu\}$  are radial,
orbital and magnetic quantum numbers, be orthonormal, i.e.,
\begin{eqnarray}
\label{orth}
&& (U_QU_{Q'}^*)=\int d^4y~U_Q(y)U_{Q'}^*(y)
=\delta_{QQ'}=\delta_{nn'}\delta_{ll'}\delta_{\{\mu\}\{\mu'\}},
\\
&& \sum\limits_QU_Q(y)U_Q^*(y')=\delta(y-y'),\nonumber
\\
&& U_Q(-y)=U_{Q}^*(y).\nonumber
\end{eqnarray}
The function $W$ can be represented by
\begin{eqnarray}
\label{Wfn}
&& W(z,-y)=\sum\limits_QW_Q(z)U_Q(-y),~~~~~~~W_Q^*(z)=W_Q(z),
\\
&& \tilde{W}(p,-y)=\sum\limits_Q\tilde{W}_Q(p)U_Q(-y),
~~~~~~~~~\tilde{W}_Q^*(p)=\tilde{W}_Q(p),\nonumber\\
&& W_Q(z)=\int{dp\over(2\pi)^4}\cdot\tilde{W}_Q(p)~
e^{-ipz}.\nonumber
\end{eqnarray}
Then we have
\begin{eqnarray}
\label{vertex}
{\cal V}(z,\stackrel{\leftrightarrow}{p}_{z})&=&
\sum\limits_QW_Q(z)\int dy~\sqrt{D_m(y)}U_Q(y)
e^{i{y\over2}\stackrel{\leftrightarrow}{p}_{z}}\\
&=&\sum\limits_QW_Q(z)
V_Q\left(\stackrel{\leftrightarrow}{p}_x\right)=(WV)_z,
\nonumber\\
V_Q\left(\stackrel{\leftrightarrow}{p}_x\right)&=&
\int dy\sqrt{D_m(y)}U_Q(y)
e^{-i{y\over2}\stackrel{\leftrightarrow}{p}_x}.\nonumber
\end{eqnarray}
In this notation one obtains
\begin{eqnarray*}
&&(A^+A)=\int dx_1\int dx_2~A^+(x_1,x_2)A(x_2,x_1)=(WW)\\
&&=\sum\limits_Q\int dz~W_Q(z)W_Q(z)
=\sum\limits_Q\int dp~\tilde{W}_Q(p)\tilde{W}_Q(p),
\end{eqnarray*}
\begin{eqnarray*}
&& {\rm tr}\ln_1\left[1+g(A^+\sqrt{D_m}){\cal D}\right]
={\rm tr}\ln_1(1+g_r(WV){\cal D}_r).
\end{eqnarray*}
The basic representation for the Green functions under
consideration becomes of the form
\begin{eqnarray}
\label{grnW}
G(x_1,x_2;x_3,x_4)&\to&G_{Q_1Q_2}(z_1,z_2)
=\int\prod\limits_Q DW_Q\cdot
W_{Q_1}(z_1)W_{Q_2}(z_2)\cdot e^{{\cal S}[W]},\\
{\cal S}[W]&=&-{1\over2}(WW)-{\rm tr}\ln_1\left[1+g_r(WV)
{\cal D}_r\right].\nonumber
\end{eqnarray}

\subsection{Particle interpretation of the quadratic term.}

Let us extract the quadratic form from ${\cal S}[W]$
\begin{eqnarray}
\label{ActW}
{\cal S}[W]&=&-{1\over2}(W[I-g_r^2\Pi]W)-
{\rm tr}\ln_2[1+g_r(WV)
{\cal D}_r],\\
&& \ln_2(1+s)=\ln(1+s)-s+{s^2\over2}.\nonumber
\end{eqnarray}
Here
\begin{eqnarray}
\label{pol1}
&& g_r^2\Pi=g_r^2{\rm tr}[V{\cal D}_rV{\cal D}_r]
\end{eqnarray}
and according to (\ref{Wfn}) and (\ref{vertex})
\begin{eqnarray*}
(Wg_r^2\Pi W)&=&\sum\limits_{QQ'}\int\!\!\int dxdx'~W_Q(x)
g_r^2\Pi_{QQ'}(x-x')W_Q(x')\\
&=&\sum\limits_Q\int dp~\tilde{W}_Q^*(p)
g_r^2\tilde{\Pi}_{QQ'}(p)\tilde{W}_Q(p).
\end{eqnarray*}
The polarization operator $g_r^2\tilde{\Pi}_{QQ'}$ looks
\begin{eqnarray}
\label{pol2}
g_r^2\Pi_{QQ'}(x-x')&=&g_r^2{\rm tr}\left\{V_Q{\cal D}_r
V_{Q'}{\cal D}_r\right\}\\
&=&g_r^2V_Q\left(\stackrel{\leftrightarrow}{p}_x\right)
{\cal D}_r(x-x')V_{Q'}\left(\stackrel{\leftrightarrow}
{p}_{x'}\right){\cal D}_r(x'-x),\nonumber\\
&=&g_r^2\int\!\!\int dydy'~U_Q(y)P(x-x';y,y')U_{Q'}^*(y'),
\nonumber\\
P(x;y,y')&=&\sqrt{D_m(y)}{\cal D}_r\left(x-{y-y'\over2}\right)
{\cal D}_r\left(x+{y-y'\over2}\right)\sqrt{D_m(y')},\nonumber\\
\tilde{P}(p;y,y')&=&\int dx~e^{ipx}P(x;y,y').\nonumber
\end{eqnarray}
In the momentum space we get
\begin{eqnarray}
\label{mtn1}
&& g_r^2\tilde{\Pi}_{QQ'}(p)=g_r^2\int{dk\over(2\pi)^4}~
V_Q(k)\tilde{{\cal D}}_r\left(k+{p\over2}\right)
\tilde{{\cal D}}_r\left(k-{p\over2}\right)V_{Q'}(k).
\end{eqnarray}
In our approximation we have
\begin{eqnarray}
\label{polapp}
g_r^2\tilde{\Pi}_{QQ'}(p)&=&
g_r^2\int{dk\over(2\pi)^4}\cdot
{V_Q(k)V_{Q'}(k)\over\left(M_r^2+\left(k+{p\over2}\right)^2
\right)
\left(M_r^2+\left(k-{p\over2}\right)^2\right)}.
\end{eqnarray}

The orthonormal system $\{U_Q(x)\}$ should be chosen so that
the polarization operator $\tilde{\Pi}_{QQ'}(p)$ should be
diagonal in radial $(n,n')$ and orbital $(l,l')$ quantum
numbers
\begin{eqnarray}
\label{polorth}
&& \tilde{\Pi}_{QQ'}(p)=\delta_{nn'}\delta_{ll'}\cdot
\tilde{\Pi}^{(nl)}_{\{\mu\}\{\mu'\}}(p)
\end{eqnarray}
The index structure of the diagonal polarization operator
$\tilde{\Pi}^{(nl)}_{\{\mu\}\{\mu'\}}(p)$ looks like
\begin{eqnarray}
&& \tilde{\Pi}^{(nl)}_{\{\mu\}\{\mu'\}}(p)
=\tilde{\Pi}^{(nl)}_{\{\mu\}\{\mu'\}}(p)
=\tilde{\Pi}^{(nl)}(p^2)\cdot\delta_{\{\mu\}\{\mu'\}}+
\sum_j\tilde{\Pi}^{(nl)}_j(p^2)\cdot t^j_{\{\mu\}\{\mu'\}}(p)
\nonumber
\end{eqnarray}
where the tensors $t^j_{\{\mu\}\{\mu'\}}(p)$ contain combinations
of the vectors $p_{\mu}p_{\mu'}$.

The diagonal quadratic form of (\ref{ActW}) gives
the equation of motion for the field
$W_Q(x)=W^{(nl)}_{\{\nu\mu_2...\mu_l\}}(x)$
\begin{eqnarray*}
&& \left[\delta_{QQ'}-g_r^2\tilde{\Pi}_{QQ'}
\left({\partial\over i\partial x}\right)\right]W_{Q'}(x)=0,
\end{eqnarray*}
or
\begin{eqnarray}
\label{eqW}
&& \left[\delta_{QQ'}-g_r^2\tilde{\Pi}_{QQ'}(p)\right]
\tilde{W}_{Q'}(p)=0.
\end{eqnarray}
The requirement that this equation on the mass shell should be
the Klein-Gordon equation gives the constraint
\begin{eqnarray}
&& {\partial\over\partial x_{\nu}}W^{(nl)}_{\nu\mu_2...\mu_l}(x)
=0 ~~~~~{\rm or}~~~~~~
p_\nu\tilde{W}^{(nl)}_{\{\nu\mu_2...\mu_l\}}(p)=0
\nonumber
\end{eqnarray}
on the mass shell. Thus, the function
$\tilde{W}^{(nl)}_{\{\mu\}}(p)$ satisfies the equation
\begin{eqnarray}
\label{eqWdg}
&& \left[1-g_r^2\tilde{\Pi}^{(nl)}(p^2)\right]
\tilde{W}^{(nl)}_{\{\mu_1...\mu_l\}}(p)=0.
\end{eqnarray}
The mass of the state with quantum numbers $Q=(nl)$ is
defined by the equation
\begin{eqnarray}
\label{eqmass}
&& 1-g_r^2\tilde{\Pi}^{(nl)}(-M_{(nl)}^2)=0.
\end{eqnarray}
Let us write
$$-1+g_r^2\tilde{\Pi}^{(nl)}(p^2)
=-Z_{(nl)}(p^2+M_{(nl)}^2)+\Sigma^{(nl)}(p^2),$$
\begin{eqnarray*}
&&\Sigma^{(nl)}(p^2)=g_r^2\tilde{\Pi}^{(nl)}_{reg}
(p^2,M_{(nl)}^2)\\
&&=g_r^2\left[\tilde{\Pi}^{(nl)}(p^2)-
\tilde{\Pi}^{(nl)}(-M_{(nl)}^2)-
\tilde{\Pi}'_{(nl)}(-M_{(nl)}^2)(p^2+M_{(nl)}^2)\right],
\end{eqnarray*}
$$ Z_{(nl)}=g_r^2\left[-\tilde{\Pi}'_{(nl)}(-M_{(nl)}^2)
\right]$$
The constant  $Z_{(nl)}$ is positive.

New field variables can be introduced as follows:
\begin{eqnarray}
\label{Wphi}
&& W_Q(x)={\varphi_Q(x)\over\sqrt{Z_{(nl)}}}.
\end{eqnarray}
The representation (\ref{grnW}) assumes of the form
\begin{eqnarray}
\label{grnphi}
&& G_{Q_0}(x-y)={1\over Z_{Q_0}}{\cal G}_{Q_0}(x-y),\\
&&{\cal G}_{Q_0}(x-y)=\int D\varphi~
\varphi_{Q_0}(x)\varphi_{Q_0}(y)~e^{{\bf S}[\phi]},\nonumber
\end{eqnarray}
where the action ${\bf S}$ looks like
\begin{eqnarray}
\label{actnphi}
&& {\bf S}[\phi]=-{1\over2}(\varphi{\cal D}^{-1}\varphi)
-{\cal I}_{int}[\varphi]
\end{eqnarray}
and the appropriate normalization should be chosen.

The kinetic term is
\begin{eqnarray}
\label{knphi}
(\varphi{\cal D}^{-1}\varphi)&=&
(\varphi\left[-\Box+M_b^2+\Sigma_b\right]\varphi)\\
&=&\int dx\sum_{\cal Q}\varphi_Q(x)
\left[-\Box+M_{(nl)}^2+\Sigma^{(nl)}(-\Box)\right]\varphi_Q(x)
\nonumber\\
&=&\int dp\sum_{\cal Q}\tilde{\varphi}_Q^+(p)
\left[p^2+M_{(nl)}^2+\Sigma^{(nl)}(p^2)\right]\tilde{\varphi}_
Q(p)\nonumber
\end{eqnarray}
and the interaction term is
\begin{eqnarray}
\label{intphi}
&& {\cal I}_{int}[\varphi]={\rm tr}\ln_2\left
[1+(h\varphi V){\cal D}\right],\\
&& (h\varphi V)=\sum_Qh_Q\varphi_QV_Q,~~~~~~~~
h_Q={1\over\sqrt{-\Pi_Q'(-M_Q^2)}}\nonumber
\end{eqnarray}

The effective dimensionless coupling constants are defined as
\begin{eqnarray}
\label{effcc}
\lambda_Q^{(eff)}&=&{h_Q^2\over16\pi M_r^2}=
{1\over 16\pi [-M_r^2\tilde{\Pi}'_{(nl)}(-M_{(nl)}^2)]}.
\end{eqnarray}

As a result, the final representation (\ref{grnphi}) can be
interpreted as a partition function of the quantum field
system of bosonic fields $\{\phi_Q\}$ which have masses $M_Q$
and are described by the nonlocal action (\ref{actnphi}).

We would like to stress that the resulting representation for the
generating functional does not contain the initial coupling
constant $g$.

All calculations with the generating functional (\ref{grnphi})
can be performed by perturbation expansions in coupling
constsnts $h_Q$. We can trust these calculations if and only if
the effective coupling constants (\ref{effcc}) are small enough:
$$\lambda_Q^{(eff)}\ll1.$$

\subsection{The orthonormal system.}

The next step is to determine the orthonormal system
(\ref{orth}). The problem is to find the spectrum and
eigenfunctions of the operator $\tilde{P}(p;y,y')$ in
(\ref{pol2}), i.e.
$$ \tilde{P}(p)U_Q=E_Q(p^2)U_Q $$
or
\begin{eqnarray}
\label{BSHerm}
&& \int dy'\tilde{{\cal P}}(p;y,y')U_Q(y',p)
=E_Q(p)U_Q(y,p),~~~~~Q=(n,l,\{\mu\}).
\end{eqnarray}
This equation can be represented in a standard form of
the Bethe-Salpeter equation in the one-boson exchange
approximation. Using the relation
$$ K_+K_-\cdot\int dx~e^{ipx}{\cal D}_r\left(x-{y-y'\over2}
\right){\cal D}_r\left(x+{y-y'\over2}\right)=\delta(y-y')$$
with
$$ K_\pm=\left[M_r^2+\left(i{\partial\over\partial y}
\pm{p\over2}\right)^2\right]$$
and introducing the functions
$$ \Psi_Q(y,p)={1\over\sqrt{D_m(y)}}\cdot U_Q(y,p)$$
we get the standard form of the Bethe-Salpeter equation
(see, for example, \cite{Itz})
\begin{eqnarray}
\label{BSeq}
&&\left[M_r^2+\left(i{\partial\over\partial y}
+{p\over2}\right)^2\right]\cdot
\left[M_r^2+\left(i{\partial\over\partial y}
-{p\over2}\right)^2\right]\Psi_Q(y,p)=g_r^2D_m(y)\Psi_Q(y,p),
\end{eqnarray}
where the spectrum is defined by the equation
\begin{eqnarray}
\label{BSE}
&& g_r^2E_Q(-M_Q^2)=1.
\end{eqnarray}
Thus the diagonalization of the operator $\tilde{P}(p;y,y')$
is equivalent to the solution of the Bethe-Salpeter equation
in one-boson exchange approximation.

It should be noted that in this formulation the functions
$\Psi_Q(y,p)$ depend on the direction of the vector
$p=(p_0,\vec{p})$ in the space ${\bf R}^4$, the standard
choice is $p_{nl}=(iM_{nl},0)$.

The main problem to use the Bethe-Salpeter basis is that the
Bethe-Salpeter equation can only be solved by numerical methods.
Even the solution obtained by Wick and Cutkosky \cite{Itz,Wick}
is reduced to the differential equation which should be
numerically computed. Our aim is to continue analytic
calculations as long as possible in order to get a visible
general picture of arising bound states in the system under
consideration. Therefore we choose a more practical way, namely,
we use an orthonormal basis that is simple enough from an
analytic point of view and is directly connected with
the problem under consideration. In this case the operators
$g_r^2\tilde{\Pi}_{QQ'}$ are not diagonal so that we should
diagonalize them. The idea consists in finding an effective basis
for diagonalization of $g_r^2\tilde{\Pi}_{QQ'}$ such that
its lowest function would provide a good qualitative description
for the eigenvelues $E_{(nl)}$ and the next two or three
functions only give a good quantitive description for those
eigenvelues.

This effective basis $\{U_Q(x)\}$ can be constructed using the
boson Green function $D_a(u)$ (\ref{prop}) with a parameter $a$
as a weight function inducing uniquely the system of orthonormal
polynomials in the space $R^4$. Thus, the full orthonormal
system of functions (\ref{orth}) can be chosen in the form
\begin{eqnarray}
\label{ortha}
&& U_Q(x,a)=i^l\sqrt{D_a(x)}a P_Q(a x).
\end{eqnarray}
Here $P_Q(u)$ are real polynomials, satisfying
$$ P_Q(-u)=(-1)^lP_Q(u),~~~~{\rm so~that}~~~~U_Q^*(x)=U_Q(-x).$$
The parameter $a$ that enters into the orthonormal system is
fixed by a variation condition formulated below.

The construction of this basis is presented in the Appendix A.

\subsection{The diagonalization procedure.}

Let us demonstrate our procedure of diagonalization of
the polarization operators (\ref{polapp}) for states $Q=(n0)$
and $Q'=(n'0)$. In the momentum space this polarization operator
looks like
\begin{eqnarray}
\label{polnn}
&& g_r^2\tilde{\Pi}_{(nn')}(p^2)
=g_r^2\int{dk\over(2\pi)^4}\cdot{V_{(n0)}(k\vert m,a)
V_{(n'0)}(k\vert m,a)\over\left(M_r^2+
\left(k+{p\over2}\right)^2\right)
\left(M_r^2+\left(k-{p\over2}\right)^2\right)}
\end{eqnarray}
where the vertex defined by (\ref{vertex}) looks in this case
like
\begin{eqnarray}
\label{vrta}
V_{(n0)}(p\vert m,a)&=&\int dx~U_{(n0)}(x)\sqrt{D_m(x)}e^{ixp}\\
&=&\int dx~\sqrt{D_m(x)D_a(x)}a P_{(n0)}(ax)e^{ixp}.\nonumber
\end{eqnarray}
One can develop
\begin{eqnarray*}
&& \sqrt{D_{m_1}(x)D_{m_2}(x)}=\sum\limits_{n=0}^\infty
\Delta^{2n}C_n\left({\Delta\over m}\right)D_m^{(n)}(x)=
D_m(x)\left[1+O\left({\Delta^2\over m^2}\right)\right]
\end{eqnarray*}
with $m={m_1+m_2\over2}$ and $\Delta={m_1-m_2\over2}$ and
$$ D_m^{(n)}=\int{dk\over(2\pi)^4}\cdot{e^{ikx}\over
(m^2+k^2)^{1+n}}.$$
We shall use the lowest approximation
\begin{eqnarray}
\label{Dappr}
&& \sqrt{D_m(x)D_a(x)}\approx D_{{m+a\over2}}(x),
\end{eqnarray}
the accuracy of which is quite acceptable for our consideration.
Then, the vertex acquires the form
\begin{eqnarray}
\label{vrtappr}
&& V_{(n0)}(p^2\vert m,a)=P_{(n0)}\left(-a^2\Box_p\right)\cdot
{\mu\over\left({m+a\over2}\right)^2+p^2}.
\end{eqnarray}
The explicit form of the polynomials $P_{(n0)}(u^2)$ is given
in the Appendix A. We get
\begin{eqnarray*}
&& g_r^2\tilde{\Pi}_{(nn')}(p^2)\\
&&={\lambda_r\over2\pi}\cdot\int\limits_0^1dt~
R_{(n)}(t)R_{(n')}(t)\cdot
{1\over b}\left[\sqrt{1+{4bt(1-t)\over(1-bt)^2}}-1\right],\\
&& R_{(n)}(t)={M_r\over t}V_{(n0)}
\left(\left.M_r^2{1-t\over t}\right\vert m,a\right).
\end{eqnarray*}

Now we formulate the variational principle which defines the
parameter $a$. The operator $\tilde{\Pi}_{(00)}$ is the
largest eigenvalue of the operator matrix $\tilde{\Pi}_{(nn')}$,
therefore the parameter $a$ can be defined by the variation
requirement
\begin{eqnarray}
\label{vara}
&& \max\limits_a~g_r^2~\tilde{\Pi}_{(00)}(-M_{(0)}^2)\\
&=&\max\limits_\eta~{\lambda_r\over2\pi}\int\limits_0^1dt~
\left({\eta\over\left({\xi+\eta\over2}\right)^2t+1-t}\right)^2
\cdot{1\over b}\left[\sqrt{1+{4bt(1-t)\over(1-bt)^2}}-1\right]
\nonumber
\end{eqnarray}
where the notion
$$ b=\left({M_{(0)}\over2M_r}\right)^2,~~~~~~\xi={m\over
M_r},~~~~~~~
\eta={a\over M_r} $$
are used. Thus, the parameter $a=a(M_{(0)},m)$ is a function
of $m$ and $M_{(0)}$. Quite a good approximation is as follows
\begin{eqnarray}
\label{amM}
&& \eta=\eta(\xi,b)\approx2-{3\over2}b+{3\over2}\xi.
\end{eqnarray}

Let us show that this orthonormal functions with the parameter
$a$ (\ref{amM}) gives quite good approximation for the
eigenvalues of the matrix $\tilde{\Pi}_{(nn')}(p^2)$. For this
aim we calculate the matrix
$$ {\cal P}_N=\left\{\tilde{\Pi}_{(nn')}(b,\xi,\eta),
~~~~(n,n'=0,...,N)\right\}$$
and their eigenvalues
$$ {\cal E}_N={\rm diag}\left\{E_0^{(N)},...,E_N^{(N)}\right\}.$$
Then we have to compare $E_j^{(N)}$ for fixed $j$ and different
$N$.

The numerical results are given in Table I.
The first case for $\xi=.5,~b=.25,~\eta=2.451$
and the second case for $\xi=.2,~b=.9,~\eta=1.22$.
One can see that for the lowest eigenvalue practically the first
lowest eigenfunction can be used, i.e. our choice of the
orthonormal
system gives quite a good accuracy.

\begin{center}
{Table I. Diagonalization of the matrix ${\cal P}_N$.}
\end{center}
\begin{center}
\begin{tabular}{|c|c|c|c|c|}
\hline
    &          &           &         &   \\
 {$N$}&{$E_0$}&{$E_1$}&{$E_2$}&{$E_3$}\\
    &          &           &         &   \\
\hline
    &          &           &         &   \\
0   & .04165  &           &         &    \\
1   & .04166  & .009941  &         &   \\
2   & .04173  & .010279  & .002755 &   \\
3   & .04175  & .010368  & .003295 & .0007482   \\
4   & .04175  & .010402  & .003546 & .0010336   \\
    &        &      &    &                  \\
\hline
    &          &           &         &   \\
0   & .1239  &           &         &    \\
1   & .1262  & .03564  &         &   \\
2   & .1262  & .03616  & .01162 &   \\
3   & .1263  & .03645  & .01298 & .003789   \\
4   & .1263  & .03655  & .01373 & .004710   \\
    &        &      &    &                  \\
\hline
\end{tabular}
\end{center}

\subsection{Applicability of S-, BS- and BNC- methods.}

Three parameters $\xi$, $\lambda_r$ and $b$ are not independent.
The standard formulation of the problem is the following:
the parameters $\xi$ and $\lambda_r$ are given and the mass of
a bound state $b$ has to be found. We reformulate this problem:
what is the region of changing $b$ for the fixed parameter $\xi$,
if the effective coupling constant $\lambda_{eff}$ is smaller
then one?

In this section we give qualitative curves
$$\frac{\Delta M}{2M}=1-{M_{(0)}(\xi)\over2M_r}
=1-\sqrt{1-b(xi)}$$
which restrict the admissible regions of $\xi$ and $b$ when the
effective coupling constant in S-, BS- and BNC-methods is smaller
then 1. These curves are presented in Fig 1. and show the
applicability of S-, BS- and BNC-methods to study the bound state
problem.

\begin{figure}[t]
\centering{\
\epsfig{figure=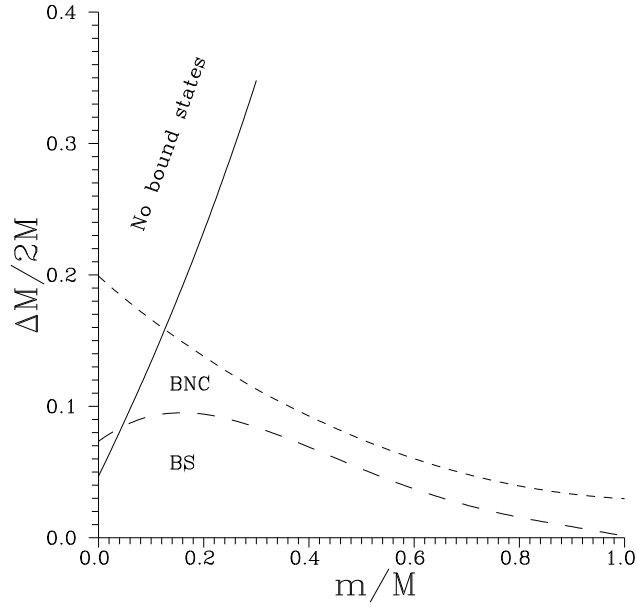,height=8cm}}
\caption{Applicability of S-, BS- and BNC- methods.}
\end{figure}

\begin{itemize}
\item The region of applicability of the nonrelativistic
Schr\"{o}dinger equation is a small vicinity near point $(0,0)$ in \\
Fig 1.
\item  The range of existence of bound states (under solid line
in Fig 1) is defined by the equation
\begin{eqnarray*}
&& 1=\lambda_{max}(\xi)\Pi_{(0)}(\xi,b)
\end{eqnarray*}
where
\begin{eqnarray*}
&& \lambda_{max}(\xi)=\max\limits_\lambda\lambda_r(\lambda,\xi)
={1\over a(\xi)},\\
&& \Pi_{(0)}(\xi,b)\\
&&=\max\limits_\eta~\int\limits_0^1{dt\over2\pi}~
\left({\eta\over\left({\xi+\eta\over2}\right)^2t+1-t}\right)^2
\cdot
{1\over b}\left[\sqrt{1+{4bt(1-t)\over(1-bt)^2}}-1\right].
\end{eqnarray*}

\item  The region of applicability of the Bethe-Salpeter method
(under dashed line in Fig 1) is defined by $\lambda<1$ and
\begin{eqnarray*}
&& \lambda_r\leq\lambda_1(\xi)=\lambda_r(1,\xi)=
{4\over(1+\sqrt{1+4a(\xi)})^2}
\end{eqnarray*}
where $b=b(\xi)$ is defined by the equation
\begin{eqnarray*}
&& 1=\lambda_1(\xi)\Pi_{(0)}(\xi,b(\xi)).
\end{eqnarray*}

\item  The rigion of applicability of the BNC-method (under
dotted line in Fig 1) is defined by the inequality
\begin{eqnarray*}
&& \lambda_0^{(eff)}(\xi,b)\leq1
\end{eqnarray*}
where the effective coupling constant is defined by
(\ref{effcc}):
\begin{eqnarray*}
\lambda_0^{(eff)}&=&{1\over 16\pi
[-M_r^2\tilde{\Pi}'_{(0)}(-M_{(0)}^2)]}={1\over\Phi(\xi,b)},\\
\Phi(\xi,b)&=&\int\limits_0^1{dt\over2\pi}~\left(
{\eta(\xi,b)\over\left({\xi+\eta(\xi,b)\over2}\right)^2t+1-t}
\right)^2\\
&\cdot&{1\over
b^2}\left\{1-{1\over\sqrt{1+{4bt(1-t)\over(1-bt)^2}}}\cdot
\left[1+{2bt(1-t)(1-3bt)\over(1-bt)^3}\right]\right\}.
\end{eqnarray*}

\end{itemize}


\subsection{Small binding energies.}

In this section we show that the BNC-method gives the correct
nonrelativistic limit in the case of the Coulomb potential.
For the Yukawa potential we show how the screening length depends
on the coupling constant and the patameter $\xi={m\over M}$.
If the binding energy is small then
$$\epsilon=1-b=1-\left({M_b\over2M}\right)^2\approx
{2M-M_b\over M}={\Delta M\over M}\ll1.$$
and one can get for (\ref{vara}) when $\epsilon$ is small
\begin{eqnarray*}
&&{1\over2\pi}\int\limits_0^1dt~\left({\eta\over\left({\xi+
\eta\over2}\right)^2t+1-t}\right)^2\cdot
{1\over b}\left[\sqrt{1+{4bt(1-t)\over(1-bt)^2}}-1\right]\\
&=&{\epsilon\over2\pi}\int\limits_0^{1/\epsilon}du~
\left({\eta\over\left({\xi+\eta\over2}\right)^2
(1-\epsilon u)+\epsilon u}
\right)^2\cdot\left[\sqrt{1+{4u(1-\epsilon
u)\over\epsilon(1+u)^2}}-1\right]\\
&\to&{1\over\epsilon^{3/2}\pi}\int\limits_0^\infty du\sqrt{u}~
\left({\eta\over\left({\xi+\eta\over2\sqrt{\epsilon}}\right)^2
+u}\right)^2
\cdot{1\over
1+u}={4\eta^2\over(\xi+\eta)(\xi+\eta+2\sqrt{\epsilon})^2}.
\end{eqnarray*}

The Coulomb potential $H={p^2\over2M_c}-{\alpha\over r}$ in
our approach is defined by the conditions $m=0$ or $\xi=0$
and $\alpha\ll1$. We get
$$ \Pi_{(0)}(0,\epsilon)=
\max\limits_\eta{4\eta\over(\eta+2\sqrt{\epsilon})^2}=
{1\over2\sqrt{\epsilon}}$$
and from
$$ 1=\lambda_r\Pi_{(0)}(0,\epsilon),~~~~{\rm or}~~~~
1={\alpha\over2\sqrt{\epsilon}},~~~~~~~~~~(\lambda_r=\alpha)$$
it follows that
$$\Delta M={\alpha^2\over2}{M_r\over2}={\alpha^2\over2}M_{cm}$$
according with the lowest energy of the nonrelativistic
Schr\"{o}dinger
equation for the Coulomb potential.

For the Yukawa potential $H={p^2\over2M_c}-
{\alpha\over r}e^{-mr}$ the critical screening length,
defined by the condition $\epsilon=0$ for the lowest
state, can be computed. For small $\xi$ this length is defined
by the equation
$$
1=\alpha\max\limits_\eta{4\eta^2\over(\xi_c+\eta)^3}={16\over2
7}\cdot
{\alpha\over\xi_c}$$
and
$$ m_c={32\over27}\alpha{M_r\over2}=1.1851...\cdot\alpha
M_{cm}.$$
or
$$ \kappa_c={m_c\over\alpha M_{cm}}={32\over27}=1.1851...$$
The numerical calculations \cite{Rog} give
$$ m_c=1.1906...\cdot\alpha M_{cm}.$$
In the case $0<\xi\leq1$ the relation between $\xi_c={m_c\over
M_r}$
and $\alpha_c$ is given by
$$\alpha_c\cdot T(\xi_c)=1~~~~~{\rm
and}~~~~~\kappa_c=2\xi_cT(\xi_c)$$
where
$$T(\xi_c)=\max\limits_\eta{1\over2\pi}\int\limits_0^1dt~
\left({\eta\over\left({\xi_c+\eta\over2}\right)^2t+1-t}\right)
^2\cdot
\left[\sqrt{1+{4t\over1-t}}-1\right].$$
The results are given in Table II. One can see that for
$\xi_c>.1$
the relativistic corrections are more then 5\%.

\begin{center}
{Table II. The critical screening length for the Yukawa
potential.}
\end{center}

\begin{center}
\begin{tabular}{|l|c|l|}
\hline
                     &            &                          \\
 $\xi_c={m_c\over M_r}$   & $\kappa_c$  & $\alpha_c={1\over
T(\xi_c)}$\\
                     &            &                           \\
\hline
    &          &              \\
0.0001   & 1.1851  &   0.00016875           \\
0.001   & 1.1846  &    0.0016888             \\
0.01    & 1.1795 &     0.01695              \\
0.1    &  1.1324   &   0.1766             \\
0.2    &  1.0861     & 0.368             \\
0.4    &  1.0079     & 0.793            \\
0.6    &   0.9436    & 1.3735             \\
0.8    &   0.8893    & 1.799             \\
1.0    &   0.8426    & 2.373             \\
\hline
\end{tabular}
\end{center}

\section{ Appendix A. Orthonormal System }

The orthonormal system $\{U_Q(x)\}$ is defined by
\begin{eqnarray*}
&& U_Q(x)=i^l\sqrt{D_a(x)}aP_Q(ax),\\
&& D_a(x)={a^2\over(2\pi)^2}\cdot{1\over
ax}K_1(ax),~~~~~~x=\sqrt{x^2}.
\end{eqnarray*}
$$ Q=(nl\{\mu\}),~~~~~~~~\{\mu\}=\mu_1,...,\mu_l.$$
\begin{eqnarray*}
&& \int d^4x~U_Q^*(x)U_{Q'}(x)
=\int d^4x~D_a(x)~aP_Q(ax)~aP_{Q'}(ax)=\delta_{QQ'},\\
\end{eqnarray*}
$$ \int d^4x=\int\limits_0^\infty drr^3\int dn,~~~~~
\int dn=\int\limits_0^{2\pi}d\phi\int\limits_0^\pi
d\theta\sin\theta\int\limits_0^\pi d\gamma\sin^2\gamma$$
$$\delta_{QQ'}=
\delta_{nn'}\delta_{ll'}\delta_{\{\mu\}\{\mu'\}},
~~~~~~\sum\limits_{\{\mu'\}}\delta_{\{\mu\}\{\mu'\}}F_{\{\mu'\}}
=F_{\{\mu\}}.$$
The polynomials are defined by
\begin{eqnarray*}
aP_Q(ax)&=&aN_{(nl)}P_{(nl)}(a^2x^2)T^{(l)}_{\{\mu\}}(ax)
=aN_{(nl)}P_{(nl)}(u^2)(\sqrt{u^2})^lT^{(l)}_{\{\mu\}}(n)
\end{eqnarray*}
$$ u=ax,~~~~~~~~ n={x\over\sqrt{x^2}}={u\over\sqrt{u^2}}.$$
The orthonormality condition looks as
\begin{eqnarray*}
&& \int dx~D_a(x)aP_Q(ax)aP_{Q'}(ax)\\
&=&{a^4\over(2\pi)^2}
\int\limits_0^\infty dxx^3~\cdot{1\over ax}K_1(ax)\int
dn~P_Q(ax)P_{Q'}(ax)\\
&=&{1\over(2\pi)^2}
\int\limits_0^\infty duu^2~K_1(u)\int dn~P_Q(u)P_{Q'}(u)\\
&=&{N_{(nl)}^2\over(2\pi)^2}\cdot 2\delta_{nn'}\cdot
{2\pi^2\over2^l(l+1)}\delta_{ll'}\cdot\delta_{\{\mu\}\{\mu'\}}=
\delta_{QQ'},~~~~~~N_{(nl)}= 2^{{l\over2}}\sqrt{l+1},
\end{eqnarray*}
\begin{eqnarray*}
&& \int dn~T^{(l)}_{\{\mu\}}(n)T^{(l')}_{\{\mu'\}}(n)=
{2\pi^2\over2^l(l+1)}\cdot\delta_{ll'}\delta_{\{\mu\}\{\mu'\}},\\
&& {1\over2}\int\limits_0^\infty ds~s^{2+2l}K_1(s)
P_{(nl)}(s^2)P_{(n'l)}(s^2)=\delta_{nn'}
\end{eqnarray*}

\subsection{Angular polynomials $T^{(l)}_{\{\mu\}}(n)$}

Polynomials $T^{(l)}_{\{\mu\}}(n)$ satisfy
\begin{itemize}
\item $T^{(l)}_{\mu_1,...,\mu_l}(n)$ is symmetric for
$\mu_i\rightleftharpoons\mu_j$,
\item $T^{(l)}_{\mu,\mu,\mu_3,...,\mu_l}(n)=0$,
\item $T^{(l)}_{\mu_1...\mu_l}(n)=
{1\over l}P(1\vert2...l)T^{(l-1)}_{\mu_2...\mu_l}(n)
-{1\over2l(l-1)}P(12\vert3...l)\delta_{\mu_1\mu_2}
T^{(l-2)}_{\mu_3...\mu_l}(n)$.
\end{itemize}
The first four polynomials are
\begin{eqnarray*}
&& T^{(0)}=1,~~~~~~~~~~T^{(1)}_{\mu}(n)=n_\mu,\\
&& T^{(2)}_{12}(n)=n_1n_2-{1\over4}\delta_{12},\\
&& T^{(3)}_{123}(n)=n_1n_2n_3-{1\over6}(n_1\delta_{23}+
n_2\delta_{31}+n_3\delta_{12}),\\
&& T^{(4)}_{1234}(n)=n_1n_2n_3n_4\\
&&-{1\over8}(n_1n_2\delta_{34}+
n_1n_3\delta_{24}+n_1n_4\delta_{23}+n_2n_3\delta_{14}+n_2n_4\d
elta_{13}+
n_3n_4\delta_{12})\\
&&+{1\over48}(\delta_{12}\delta_{34}+
\delta_{13}\delta_{24}+\delta_{14}\delta_{23})\\
\end{eqnarray*}
where the condensed notation $n_j=n_{\mu_j}$ and
$\delta_{ij}=\delta_{\mu_i\mu_j}$ are introduced.

The normalization of the angular polynomials is defined by
\begin{eqnarray*}
&& \sum\limits_{\{\mu\}}T^{(l)}_{\mu_1...\mu_l}(n_1)
T^{(l)}_{\mu_1...\mu_l}(n_2)={1\over2^l}C_l^1(t),
~~~~~~~~t=(n_1n_2),\\
&& C_l^1(t)=\sqrt{{\pi\over2}}p_l(t),\\
&& \int\limits_{-1}^1dt~\sqrt{1-t^2}\cdot p_l(t)p_{l'}(t)=
\int\limits_0^\pi
d\gamma~\sin^2\gamma~p_l(\cos\gamma)p_{l'}(\cos\gamma)=
\delta_{ll'}
\end{eqnarray*}
where $C^1_l(t)$ are the Gegenbauer polynomials. In addition one can get
\begin{eqnarray*}
&& \int dn~(T^{(l)}_{1...l}(n_1)T^{(l)}_{1...l}(n))\cdot
(T^{(l)}_{1...l}(n)T^{(l)}_{1...l}(n_2))\\
&&={2\pi^2\over2^l(l+1)}(T^{(l)}_{1...l}(n_1)T^{(l)}_{1...l}(n
_2))=
{2\pi^2\over2^{2l}(l+1)}C_l^1((n_1n_2))
\end{eqnarray*}

\subsection{Radial functions}

The radial functions are defined in a standard way (see,
for example, \cite{Ryzh}). Let us define
\begin{eqnarray*}
c(n)&=&{1\over2}\int\limits_0^\infty
du~u^{2+2n}K_1(u)=4^n(n+1)!n!,\\
C_n(l)&=&\left\vert
\begin{array}{cccc}
c(l)   & c(l+1) & ... & c(l+n) \\
c(l+1) & c(l+2) & ... & c(l+n+1) \\
...    & ...    & ... & ... \\
c(l+n) & c(l+n+1) & ... & c(l+2n)
\end{array}
\right\vert.
\end{eqnarray*}
The radial polinomials read:
\begin{eqnarray*}
&& P_{(nl)}(u^2)={1\over\sqrt{D_{n-1}(l)D_n(l)}}\\
&& \cdot\left\vert
\begin{array}{ccccc}
c(l)   & c(l+1) & c(l+2) &... & c(l+n) \\
c(l+1) & c(l+2) & c(l+3) &... & c(l+n+1) \\
...    & ...    & ... & ... & ...\\
c(l+n-1) & c(l+n) & c(l+n+1) & ... & c(l+2n-1)\\
1      &  u^2    & (u^2)^2 & ... & (u^2)^n
\end{array}
\right\vert
\end{eqnarray*}
where $ D_n(l)={\rm det}~C_n(l)$.

The orthonormality condition is as follows:
\begin{eqnarray*}
&&{1\over2}\int\limits_0^\infty
du~u^{2+2l}K_1(u)P_{(nl)}(u^2)P_{(n'l)}(u^2)
=\delta_{nn'}.
\end{eqnarray*}
The first four polynoms are
\begin{eqnarray*}
&& aP_{(00)}=a, \\
&& aP_{(10)}(ax)={a\over\sqrt{2}}\left[-1+{a^2x^2\over8}\right],
\\
&& aP_{(20)}(ax)={a\over\sqrt{26}}\left[3-{5a^2x^2\over8}+
{(a^2x^2)^2\over96}\right], \\
&& aP_{(30)}(ax)={a\over\sqrt{4303}}\left[-34+{37a^2x^2\over4}-
{53(a^2x^2)^2\over192}+{13(a^2x^2)^3\over9216}\right].
\end{eqnarray*}

\subsection{Vertex functions}

The vertex function
\begin{eqnarray*}
V_Q(k)&=&i^l\int dx~\sqrt{D_m(x)D_a(x)}~aP_Q(ax)e^{-ikx}
\end{eqnarray*}
in the approximation
\begin{eqnarray*}
&& \sqrt{D_m(x)D_a(x)}\approx D_{{m+a\over2}}(x)
\end{eqnarray*}
looks as
\begin{eqnarray*}
V_Q(k)&\approx&i^l\int dx~D_{{m+\mu\over2}}(x)aP_Q(ax)
e^{-ipx}=i^lP_Q\left(ia{\partial\over\partial k}\right)\cdot
{a\over \left({m+a\over2}\right)^2+k^2}.
\end{eqnarray*}
The following equality is useful:
\begin{eqnarray*}
&&\left({\partial^2\over\partial k^2}\right)^s\cdot{1\over
M^2+k^2}
=\Box_k^s\cdot{1\over M^2+k^2}\\
&&=\sum\limits_{j=0}^s(-1)^{s-j}
\cdot{4^ss!(s+1)!(s+j)!\over j!(j+1)!(s-j)!}
\cdot{(k^2)^j\over(M^2+k^2)^{s+j+1}}
\end{eqnarray*}


\begin{references}
\bibitem{Itz}
C.Itzykson and J.-B.Zuber, {\it Quantum Field Theory},
McGraw-Hill Book Company, N.Y., 1980.
\bibitem{Sapir}
J.R.Sapirstein and D.R.Yennie,
{\it Theory of Hydrogenic Bound States} in {\it Quantum
Electrodynamics}, Ed. T.Kinishita, Advanced Series on
Directions in High Energy Physics, vol 7,
World Scientific, 1999.
\bibitem{Grein}
W.Greiner and J.Reinhart, {\it Quantum Electrodinamics},
Springer-Verlag, Berlin, N.Y., 1992.
\bibitem{Kapt}
S.M.Dorkin, L.P.Kapteri and S.S.Semikh,
Yad. Fiz. {\bf 60}, 1784 (1997).
\bibitem{Luch}
W.Lucha, {\it et~~al}, Phys. Rep. {\bf 200}, N4 (1991);
C.Quigg and J.L.Rosner, Phys. Rep. {\bf 56}, 167 (1976).
\bibitem{Griff}
D.Griffits, {\it Introduction to
Elementary Particles}, Harper \& Row.Publishers, NY, 1987.
\bibitem{Wick}
G.C.Wick, Phys. Rev. {\bf 96}, 1124 (1954);
R.E.Cutkosky, Phys. Rev. {\bf 96}, 1135 (1954).
\bibitem{Hay}
K.Hayashi, {\it et~~al}, Fort.Phys. {\bf 15}, 625 (1967).
\bibitem{Nak}
N.Nakanishi, Suppl. of Prog. Theor. Phys. {\bf 43}, 1 (1969).
\bibitem{Efiv}
G.V.Efimov and M.A.Ivanov, {\it The Quark Confinement Model
of Hadrons}, IOP Publishing Ltd, Bristol and Philadelphia, 1993.
\bibitem{Rob}
R.T.Cahil and C.D.Roberts, Phys. Rev. {\bf D32}, 2419 (1985);
C.D.Roberts, {\it et~~al}, Phys. Rev. {\bf D49}, 125 (1994).
\bibitem{Kop}
P.Kopietz, {\it Bosonization of Interacting Fermions in
Arbitrary Dimensions}, Lecture Notes in Physics, {\bf m48},
Springer-Verlag, Berlin Heidelberg N.Y., 1996.
\bibitem{Dar}
J.W.Darewych, Can. J. Phys. {\bf 76}, 523 (1998).
\bibitem{Efned}
G.V.Efimov and S.N.Nedelko, Phys. Rev. {\bf D51}, 176 (1995);
Phys. Rev. {\bf D54}, 4483 (1996).
\bibitem{Bog}
N.N.Bogoliubov, D.V.Shirkov, {\it Introduction to the Theory
of Quantized Fields}, Interscience Publishers, Ltd, N.Y., 1959.
\bibitem{Diney}
M.Dineykhan, {\it et~~al}, {\it Oscillator Representation in
Quantum Physics}, {\bf m24}, Springer-Verlag, Berlin Heidelberg
N.Y., 1995.
\bibitem{Feyn}
R.P.Feynman and A.R.Hibbs, {\it Quantum Mechanics and Path
Integrals}, McGraw-Hill Book Company, N.Y., 1965.
\bibitem{Rog}
F.J.Rogers, {\it et~~al}, Phys. Rev. {\bf A1}, 1577 (1970).
\bibitem{Ryzh}
I.S.Gradshtain and I.M.Ryzhik, {\it Tables of Integrals, Series
and Products}, Academic Press, New York, 1980.
\end{references}
\end{document}